\documentclass[twocolumn,showpacs,preprintnumbers,amsmath,amssymb]{revtex4}

\usepackage{graphicx}% Include figure files
\usepackage{dcolumn}% Align table columns on decimal point
\usepackage{bm}% bold math
\begin{document} 

\title{Electronic Structure of NiO: Antiferromagnetic Transition and 
Photoelectron Spectra in the Ordered Phase}
\author{R. Eder}
\affiliation{Karlsruhe Institute of Technology,
Institute for Solid State Physics, 76021 Karlsruhe, Germany}
\date{\today}

\begin{abstract}
The thermodynamics of the antiferromagnetic ordering transition in NiO and
the photoelectron spectra in the antiferromagnetic phase are studied by the 
Variational Cluster Approximation.  Using realistic Racah parameters 
to describe the Coulomb interaction in the Ni 3d shell and a Slater-Koster 
parameter $(pd\sigma)$ which is slightly ($10\%$) increased over the
band structure estimate the calculated N\'eel temperature is 481 Kelvin 
(experimental value: 523 Kelvin). The magnetic susceptibility above $T_N$ has 
Curie-Weiss form. A significant contribution to the stabilization of the 
antiferromagnetic phase comes from electron hopping between oxygen which would 
be missed in theories that consider superexchange along a single bond only.
The single particle spectral function in the ordered phase is in
good agreement with experiment, in particular a number of dispersionless 
bands which are not reproduced by most calculations are obtained correctly. 
These flat bands are shown to be direct experimental evidence for a 
dispersionless electronic self-energy with several poles in the energy range 
of the valence band which originate from the multiplets of the Ni$^{3+}$ ion. 
Small but possibly experimentally detectable changes of the photoelectron 
spectra with temperature are discussed, in particular a widening of the 
insulating gap in the paramagnetic phase by approximately $10\%$ is predicted.
\end{abstract} 

\maketitle
%%%%%%%%%%%%%%%%%%%%%%%%%%%%%%%%%%%%%%%
\section{Introduction}
%%%%%%%%%%%%%%%%%%%%%%%%%%%%%%%%%%%%%%%
Nickel Oxide has received attention over several 
decades because it is the prototype of a correlated insulator.
As early as 1937 this material was cited as a counterexample 
to the Bloch theory of solids\cite{deBoer}: assuming strong ionicity,
oxygen will be O$^{2-}$ leaving Nickel to be Ni$^{2+}$ or [Ar]3d$^8$.
This means that $4$ electrons/spin direction have to be distributed over
five Ni 3d bands which could result in an insulating ground state
only if one of the five bands were splitt off from the others over the entire 
Brillouin zone. This is impossible, however, because at $\Gamma$ the five Ni 3d
bands converge into one 3-fold degenerate ($t_{2g}$)
and one 2-fold degenerate ($e_g$) level.
Band structure calculations for the paramagnetic phase of NiO\cite{Mattheiss}
confirm this, showing three O 2p derived bands
well below the Fermi energy and a group of five Ni 3d bands which are
intersected by the Fermi energy.\\
This simple picture is modified in that NiO undergoes an antiferromagnetic 
ordering transition at $T_N=523$ Kelvin.
Thereby Ni-ions in planes perpendicular to
$(1,1,1)$ allign their magnetic moments parallel to each other,
with the ordered moment in successive planes being antiparallel
(type II antiferromagnetism). This structure is consistent with
the Goodenough-Kanamori rules because all $180^o$ Ni-O-Ni bonds are
antiferromagnetic 
in this way.
Band structure calculations within the framework of Density Functional Theory
(DFT) for the antiferromagnetic phase reproduce the insulating ground state, 
but the band gap is only $G\approx 0.3\;eV$\cite{Terakura} whereas the
experimental value is $G\approx 4.3\;eV$\cite{SawatzkyAllen}.
It should be noted, however, that DFT does
indeed give a rather accurate estimate of $G=4.1\;eV$ for the 
single-particle gap in NiO\cite{NormanFreeman}
if one does not use the band structure of Kohn-Sham eigenvalues
- which have no true physical significance anyway -
but calculates the ground state energy $E_0$ of finite clusters
as a function of electron number $N$ and uses 
$G=E_0(N+1) + E_0(N-1) - 2E_0(N)$.
The crucial point is, however, that NiO remains an insulator even 
above $T_N$ so that the insulating 
nature of NiO cannot be explained by antiferromagnetic ordering.\\
There is general agreement by now that the true origin
of the insulating nature of NiO is the strong Coulomb interaction between
electrons in the Ni 3d shell so that an adequate description requires
a more accurate treatement of the electron-electron interaction.
Accordingly, a wide variety of methods for treating interacting electrons 
have been applied to NiO over the years.
Amongst others there are calculations using the self-interaction
corrected density functional theory\cite{SvaGu,Szotek},
the LDA+U formalism\cite{Czyzyk,Bengone} and the 
GW approximation\cite{Ary,Massi} which more recently, was also
combined with the LDA+U formalism\cite{Kobayashi_etal}.
NiO was also treated in the framework of the three-body scattering 
formalism\cite{Manghi,Iga} and dynamical mean-field theory (DMFT),
both for the paramagnetic\cite{Kunesetal,Kunesetal_band,yinetal}
and antiferromagnetic\cite{MiuraFujiwara} phase.
Moreover it was shown recently that within DFT
the agreement between calculated and 
measured band gap for antiferromagnetic NiO
is improved considerably if a more accurate density functional
is used\cite{GillenRobertson}.\\
%%%%%%%%%%%%%%%%%%%%%%%%%%%LDA%%%%%%%%%%%%%%%%%%%%%%%%%%%%%%%%%%%%%%
%%%%%%%%%%%%%%%%%%%%%%Andere Methoden%%%%%%%%%%%%%%%%%%%%%%%%%%%%%%%
%%%%%%%%%%%%%%%%%%%%%%%%%%%%%%%%%%%%%%%%%%%%%%%%%%%%%%%%%%%%%%%%%%%%
A quite different - but very successful - approach 
was initiated by Fujimori and Minami\cite{c1}.
These authors showed that good agreement between theory and
experiment could be obtained for angle-integrated
valence band photoemission spectra
if one focused on local physics by considering an octahedron-shaped
NiO$_6^{10-}$ cluster comprising of a single Ni-ion and its six nearest 
neighbor oxygen ions. The eigenstates and eigenenergies of
such a finite cluster can be calculated exactly by 
the configuration interaction (CI)
(or exact diagonalization) method and the single-particle
spectral function be obtained from its Lehmann representation.
The CI method was subsequently applied to the calculation of
the angle-integrated valence band photoemission spectra of a number of 
transition metal compounds\cite{c2,c3,c4,c5,c6,c7,c8,c9,c10}
and was extended to simulate
X-ray absorption spectra\cite{x0,x1,x2,x3,x4,x5,x6,x7,x8}.
In all cases the agreement with
experiment is excellent and in the case of  X-ray absorption 
spectroscopy the comparison of simulated and measured spectra
by now is in fact becoming a routine tool for determining the
valence and spin state of transition metal ions in solids\cite{xr1,xr2}.\\
The reason for the success of the cluster method is 
that it takes into account the full Coulomb interaction between 
electrons in the transition metal 3d shell in the framework of atomic 
multiplet theory\cite{Slater,Griffith,Sugano}.
Introducing the compound index $\nu=(n,l,m,\sigma)$
(where $n=3$ and $l=2$ for a 3d-shell)
the Coulomb interaction between electrons in atomic shells
can be written as\cite{Slater,Griffith,Sugano}
\begin{eqnarray}
H_1&=&\frac{1}{2}\;\sum_{i,j,k,l} V(\nu_i,\nu_j,\nu_k,\nu_l)\;
c_{\nu_i}^\dagger c_{\nu_j}^\dagger  c_{\nu_l}^{} c_{\nu_k}^{}, \nonumber \\
&&\nonumber\\
V(\nu_1,\nu_2,\nu_3,\nu_4)&=&\delta_{\sigma_1,\sigma_4}\;
\delta_{\sigma_2,\sigma_3}\;\delta_{m_1+m_2,m_3+m_4}\;\nonumber\\
&&\nonumber\\
&& \sum_{k}\;
F^k \;c^k(l_1m_1;l_4m_4)\;c^k(l_3m_3;l_2m_2).\nonumber \\
\label{h1}
\end{eqnarray}
Here the $c^k(lm_3;lm_2)$ denote Gaunt coefficients the
$F^k$ Slater integrals and for a d-shell the multipole
index $k\in\{0,2,4\}$.
The Hamiltonian (\ref{h1}) was derived orginally to
explain the line spectra of atoms and ions in vacuum,
see Ref. \cite{Slater} for an extensive list of examples.
There are various simplified expressions in the literature\cite{Kanamori}
where the Hamiltonian (\ref{h1}) is approximated in terms of
Hubbard-$U$ and Hund's rule $J$ and sometimes additional parameters
(see Ref. \cite{Marel} for the relation between 
Hubbard-$U$ and Hund's rule $J$ and the Slater integrals)
but the Hamiltonian (\ref{h1}) is the only one that can actually
be derived from first principles and gives correct results for free ions
(an instructive comparison of the eigenvalue 
spectra of the original  Hamiltonian (\ref{h1}) and various simplified
versions was given by Haverkort\cite{maurits_thesis}).
$H_1$ contains diagonal terms such as
\begin{eqnarray}
\left(\;
V(\nu_1,\nu_2,\nu_1,\nu_2)
-V(\nu_1,\nu_2,\nu_2,\nu_1) \;
\;\right)
n_{\nu_1}\; n_{\nu_2}
\label{diagonal}
\end{eqnarray}
but also off-diagonal terms where all four $\nu_i$ in
(\ref{h1}) are pairwise
different. The off-diagonal matrix elements are frequently discarded in 
DMFT calculations\cite{Kunesetal,Kunesetal_band}
because they exchange electrons and thus exacerbate the minus-sign
problem in quantum Monte-Carlo calculations. On the other hand
the matrix elements of these terms are of the same order of magnitude
(namely proportional to the Slater integrals $F^2$ and $F^4$) as the
differences between the various diagonal matrix elements 
in (\ref{diagonal}) so that there is no justification for discarding them 
but keeping different diagonal matrix elements. \\
As will be seen below the Variational Cluster Approximation (VCA)
proposed by Potthoff\cite{p1,p2,p3} allows to extend  the scope of the 
CI method of Fujimori and Minami once more, to the calculation of 
thermodynamical quantities and band structures for strongly correlated 
electron systems. Since the VCA is based on exact diagonalization and 
therefore free from the minus-sign problem the full Coulomb Hamiltonian
(\ref{h1}) including the off-diagonal matrix elements can be included.
As could have been expected on the basis of the considerable
success of the CI method in reproducing experimental spectra
\cite{c1,c2,c3,c4,c5,c6,c7,c8,c9,c10,x0,x1,x2,x3,x4,x5,x6,x7,x8}
the VCA achieves good agreement with experiment.
%%%%%%%%%%%%%%%%%%%%%%%%%%%%%%%%%%%%%%%%%%%%
\section{Hamiltonian and Method of Calculation}
%%%%%%%%%%%%%%%%%%%%%%%%%%%%%%%%%%%%%%%%%%%%
The method of calculation has been described in detail in Ref. \cite{vca_1}
so we give only a brief description. The Hamiltonian describing the NiO 
lattice is
\begin{eqnarray}
H&=&H_0 + H_1 \nonumber \\
H_0 &=&
\sum_{i\alpha\sigma} 
\epsilon_{\alpha} \;d_{i\alpha\sigma}^\dagger d_{i\alpha\sigma}^{}
%+ \sum_{j\beta\sigma}
% \epsilon_{\beta} \;p_{j\beta\sigma}^\dagger p_{j\beta\sigma}^{}
+ H_{pd} + H_{pp}+ H_{dd}
\nonumber \\
 H_{pd} &=& \sum_{i\alpha,j\beta}\sum_\sigma\;(t_{i\alpha,j\beta}
\;d_{i\alpha\sigma}^\dagger p_{j\beta\sigma}^{} + H.c.)
%\nonumber \\
%&&+ \sum_{j\beta,j'\beta'}\sum_\sigma\;(t_{j\beta,j'\beta'}
%\;p_{j\beta\sigma}^\dagger p_{j'\beta'\sigma}^{} + H.c.)
%\nonumber \\
%&&+ \sum_{i\alpha,i'\alpha'}\sum_\sigma\;(t_{i\alpha,i'\alpha'}
%\;d_{i\alpha\sigma}^\dagger d_{i'\alpha'\sigma}^{} + H.c.),
\label{h0}
\end{eqnarray}
where e.g. $d_{i\alpha\sigma}^\dagger$ creates an electron with $z$-spin
$\sigma$ in the
Ni 3d orbital $\alpha\in\{xy, xz,\dots,3z^2-r^2\}$ at the Ni-site $i$
whereas $p_{j\beta\sigma}^\dagger$ creates an electron in the O 2p orbital
$\beta \in \{x,y,z\}$ at the O-site $j$. The energy of the
O 2p orbitals is the zero of energy.
The terms $H_{pp}$ and $H_{dd}$ describe hopping between two
O 2p orbitals or two Ni 3d orbitals, respectively, and their form is
self-evident.
The parameters in this Hamiltonian have been obtained from a fit to an
LDA band structure and are listed in Table I of Ref. \cite{vca_1}. 
One noteworthy detail is
that the energies $\epsilon_{\alpha}$ have to be subject to the
`double-counting correction': $\epsilon_{\alpha} \rightarrow 
\epsilon_{\alpha} - nU$ with $U$ the Hubbard $U$ - see Ref. \cite{vca_1} 
for a detailed discussion. 
The interaction Hamiltonian $H_1$ has the form (\ref{h1}) for each Ni 3d shell,
the Racah parameters were $A=7\;eV$, $B=0.13\;eV$ $C=0.6\;eV$, 
resulting in the Slater integrals $F^0=7.84\;eV$, $F^2=10.57\;eV$ 
and $F^4=7.56\;eV$.
An important detail is that there is a nonvanishing interaction only
between Ni 3d orbitals in the same Ni ion.\\
For a multiband system such as (\ref{h0}) the imaginary time
Green's function ${\bf G}({\bf k},i\omega_\nu)$ and self-energy 
${\bf \Sigma}({\bf k},i\omega_\nu)$ are matrices of dimension
$2n_{orb}\times 2n_{orb}$ with $n_{orb}$ the number of orbitals/unit cell.
In the following we will often omit the ${\bf k},i\omega_\nu$ argument
on these quantities for brevity.\\
The starting point for the VCA is an expression for the Grand Canonical
Potential of an interacting Fermi system derived by 
Luttinger and Ward\cite{LuttingerWard}
\begin{eqnarray}
\Omega = -
\frac{1}{\beta}\;\sum_{{\bf k},\nu}\; e^{i\omega_\nu 0^+}\left(
\ln\mbox{det}\left(-{\bf G}^{-1}\right)
+ \mbox{tr}\;{\bf \Sigma}{\bf G}\right) +\Phi[{\bf G}].
\nonumber \\
%&&\;\;\;\;\;\;\;\;\;\;\;\;\;\;\;\;\;\;\;\;\;+\Phi[{\bf G}].
\label{luttinger_ward}
\end{eqnarray}
Here $\Phi[{\bf G}]$ denotes the so-called
Luttinger-Ward functional which was defined orginally\cite{LuttingerWard}
as a sum over infinitely many closed, connected, skeleton diagrams with the
noninteracting Green's function ${\bf G}_0$ replaced by the
argument of the functional, ${\bf G}$. A nonperturbative construction 
of $\Phi[{\bf G}]$ has been given by 
Potthoff\cite{Nonperturbative,review}. 
In their proof of (\ref{luttinger_ward})
Luttinger and Ward derived two important results:
first, $\Phi[{\bf G}]$ is the generating functional of the self-energy
\begin{eqnarray}
\frac{\partial \Phi[{\bf G}]}{\partial G_{\alpha\beta}({\bf k},i\omega_\nu)}
=-\frac{1}{\beta}\;\Sigma_{\beta\alpha}({\bf k},i\omega_\nu)
\label{generating}
\end{eqnarray}
and, second, $\Omega$ is stationary under variations of 
${\bf \Sigma}$
\begin{eqnarray}
\frac{\partial \Omega}{\partial \Sigma_{\alpha\beta}({\bf k},i\omega_\nu)}
=0.
\label{stationary}
\end{eqnarray}
The first of these equations can be used\cite{p1,p2} 
to define the Legendre transform $F[{\bm \Sigma}]$ of $\Phi[{\bf G}]$ via
\begin{eqnarray*}
F[{\bf \Sigma}] &=& \Phi[ {\bf G}[{\bf \Sigma}]] - \frac{1}{\beta}
\sum_{{\bf k},\nu}\;\mbox{tr}\;{\bf \Sigma}\;{\bf G}.
\end{eqnarray*}
Introducing the noninteracting Green's function ${\bf G}_0$, 
$\Omega$ thus can be expressed as a functional of
${\bf \Sigma}$:
\begin{eqnarray}
\Omega &=& -\frac{1}{\beta}
\;\sum_{{\bf k},\nu}\; e^{i\omega_\nu 0^+}\left[
\ln\;\mbox{det}
\left( -{\bf G}_0^{-1} + {\bf \Sigma}\right)
\;\right] + F[{\bf \Sigma}]\nonumber \\
\label{basic}
\end{eqnarray}
which is known to be stationary at the exact ${\bf \Sigma}(\omega)$
by virtue of
(\ref{stationary}). The problem one faces in the practical application of 
this stationarity principle is that no explicit 
functional form of $F[{\bf \Sigma}]$ is known.\\
In the framework of the VCA this problem is circumvented 
as follows\cite{p1,p2,p3}: first, we note that $\Phi[{\bf G}]$ 
involves only the interaction part $H_1$ of the Hamiltonian
(via the interaction lines in the skeleton diagrams) and the Green's function 
${\bf G}$ (via the Green's function lines) - the latter, however,
is the argument of the functional. This implies that the functional 
$\Phi[{\bf G}]$ 
and its Legendre transform $F[{\bf \Sigma}]$ are the same for 
any two systems with the same interaction part $H_1$
(Potthoff has derived this property without making any reference to 
diagrams\cite{Nonperturbative,review}).\\
In the application to NiO we accordingly consider two systems:
System I is the original NiO lattice described by the Hamiltonian (\ref{h0})
whereas System II - termed the reference system by Potthoff\cite{p1,p2,p3} -
is an array of clusters, each of which consists of the five Ni 3d orbitals 
of one Ni ion of the original NiO lattice plus
five Ligands or bath sites\cite{p1,p2} which hybridize
with these. The single-particle Hamiltonian of such a cluster is
\begin{eqnarray}
\tilde{H}_0 &=&\sum_{\alpha,\sigma}(
\epsilon_{d}(\alpha) \;d_{\alpha,\sigma}^\dagger d_{\alpha,\sigma}^{}
+\epsilon_{L}(\alpha)\; l_{\alpha,\sigma}^\dagger l_{\alpha,\sigma}^{})
\nonumber \\
&& + \sum_{\alpha,\sigma}(
V(\alpha) \; d_{\alpha,\sigma}^\dagger l_{\alpha,\sigma}^{} + H.c.),
\label{clukin}
\end{eqnarray} 
where $\alpha \in\{xy,xz\dots3z^2-r^2\}$ whereas
the interaction part $H_1$ for each cluster is again given
by (\ref{h1}).
In the CI method by Fujimori and Minami the Ligand $l_\alpha$
would be the linear combination of O 2p orbitals
on the $6$ oxygen ions surrounding the Ni ion under
consideration which hybridizes with the d-orbital
$d_\alpha$. In the case of the VCA the Ligands
are purely mathematical objects which have no counterpart in the
physical system and which are introduced solely
for the purpose of constructing self-energies. Accordingly, there are
{\em no} terms coupling the clusters centered on neighboring
Ni ions in system II which therefore consists of completely disconnected finite 
clusters. The crucial point is, that since the interaction parts of
systems I and II are identical by construction they have the same 
Luttinger-Ward functional $F[{\bf \Sigma}]$. 
Since the individual clusters of system II are relativey small -
they comprise 10 orbitals/spin direction - they can be treated by
exact diagonalization and we can obtain all eigenstates of
$H-\mu N$ within $\approx 20 k_B T$ above the minimum value.
Using these the Grand Potential $\tilde{\Omega}$ can be evaluated numerically
(quantities with $\tilde{}$ refer to a cluster in the following) 
and the full Green's function $\tilde{\bf G}(\omega)$ be calculated
(e.g. by using the Lanczos algorithm). Next, $\tilde{\bf G}(\omega)$ can be
inverted numerically for each $\omega$ and the self-energy
$\tilde{\bf \Sigma}(\omega)$ be extracted. 
Thereby we have in real-space representation
$\tilde{\Sigma}_{\alpha\beta}(i,j,\omega)= 
\tilde{\Sigma}_{\alpha\beta}(\omega)\; \delta_{ij}$ 
where $i,j$ are the indizes of the individual disconnected clusters and
moreover $\tilde{\Sigma}_{\alpha\beta}(\omega)\ne 0$ only if both
indizes $\alpha$ and $\beta$ refer to Ni 3d orbitals.
The resulting self-energy thus is ${\bf k}$-independent and
bears no more reference to the ficticious Ligands.\\
Using $\tilde{\Omega}$ and $\tilde{\bf \Sigma}(\omega)$
the equation (\ref{basic}) - now applied to
a single cluster - can be reverted to obtain the numerical
value of $F[\tilde{\bf \Sigma}]$ for the 
self-energy $\tilde{\bf \Sigma}(\omega)$. By taking the disgression
to the reference system of clusters it is thus possible to generate self-energies
for which the exact numerical value of the Luttinger-Ward functional is known.
Next, these self-energies are used
as `trial self-energies' for the lattice i.e. we approximate
\begin{eqnarray}
\Omega &\approx& -\frac{1}{\beta}
\;\sum_{{\bf k},\nu}\; e^{i\omega_\nu 0^+}\left[\ln\;\mbox{det}
\left( -{\bf G}_0^{-1} + 
\tilde{\bf \Sigma}\;\right)
\;\right] \nonumber + N F[\tilde{\bf \Sigma}] \nonumber \\
\label{trial}
\end{eqnarray}
where ${\bf G}_0$ 
now is the noninteracting Green's function
of the physical NiO lattice and $N$ the number of
Ni-sites in this.\\
The variation of $\tilde{\bf \Sigma}$ is performed by varying the
single particle parameters $\lambda_i$ of the cluster single-particle
Hamiltonian (\ref{clukin}), that means $\epsilon_d(\alpha)$,
$\epsilon_{L}(\alpha)$ and $V(\alpha)$. These parameters
are not determined as yet because the only requirement
for the equality of the Luttinger-Ward functionals of the two systems
was that the {\em interaction parts} $H_1$ be identical.
In this way the approximate $\Omega$ (\ref{trial})
becomes a function of the
$\lambda_i$, $\Omega=\Omega(\lambda_1,\dots\lambda_n)$
and the stationarity condition (\ref{stationary}) is replaced 
by a condition on the $\lambda_i$:
\begin{eqnarray}
\frac{\partial \Omega}{\partial \lambda_i}=0.
\label{vca}
\end{eqnarray}
The physical interpretation would be that the VCA amounts to
seeking the best approximation to the true self-energy of the NiO lattice 
amongst the `cluster representable' ones. Since its
invention by Potthoff the VCA has been applied 
to study the Hubbard model in various 
dimensions\cite{h1,h2,h3,h4,h5,h6,h7,h8,h9,h10,h11}
models for 3d transiton metal compounds\cite{r1,r2,r3,vca_lacoo3,r4}
and interacting Bosons\cite{b1,b2}.\\
For the present application to NiO and
in the paramagnetic case cubic
symmetry reduces the number of parameters $\lambda_i$ to be varied
to only six: for each $\alpha$
the Hamiltonian (\ref{clukin}) contains $3$ parameters
and there is one such set for the $e_g$ orbitals and 
one for the $t_{2g}$ orbitals.
The equation system (\ref{vca}) is solved by the Newton method,
see Refs. \cite{vca_1}, \cite{vca_lacoo3} for details.
The proposal of Balzer and Potthoff\cite{h6} to use rotated and rescaled
coordiante axis for the calculation of the derivatives of
$\Omega$ with respect to the $\lambda_i$ turned out to be
of crucial importance for successful Newton iterations.\\
The paramagnetic phase of NiO was studied in some detail
in the preceeding paper Ref. \cite{vca_1}.
There are only a few differences as compared to this study:
first, a reduced value
of the Racah-parameter $A=7.0\;eV$ ($A=8.25\;eV$ in
Ref. \cite{vca_1}) and consequently a readjustment of the Ni 3d-orbital energy 
$\epsilon_d=-52\;eV$
(whereas $\epsilon_d=-62\;eV$ in Ref. \cite{vca_1} ) 
to account for the different
double-counting correction. Moreover, the Ni 3d-to-O 2p hopping parameter
$(pd\sigma)$ was increased by $10\%$ to $-1.4178\;eV$.\\
Due to improved computer power 
it was now moreover possible to optimize all relevant single particle
parameters of the octahedral cluster.
Thereby it turned out that $V(t_{2g})=0$ is a stationary point
irrespective of the values of the other parameters.
The energy of the $t_{2g}$-like Ligand, $\epsilon_L(t_{2g})$
then is irrelvant so that only the four parameters
$\epsilon_{d}(e_g)$, $\epsilon_{L}(e_g)$, $V(e_g)$ and
$\epsilon_{d}(t_{2g})$ remain to be solved for.
The resulting paramagnetic solution, however, is stationary with respect to
all six possible parameters.
Some results for the paramagnetic phase will be presented later
in comparison to the antiferromagnetic one.
%%%%%%%%%%%%%%%%%%%%%%%%%%%%%%%%%%%%%%%%%%%%
\section{Magnetic Susceptibilities and Antiferromagnetic Transition}
%%%%%%%%%%%%%%%%%%%%%%%%%%%%%%%%%%%%%%%%%%%%
We discuss the staggered and uniform magnetic susceptibility.
Within the VCA the Grand Canonical Potential may be thought of
as being expressed as a function of a number of parameters
\begin{equation}
\Omega = \Omega(\zeta_1,\dots \zeta_m,\lambda_1,\dots,\lambda_n),
\label{Omegafunction}
\end{equation}
were the $\zeta_i$ are the parameters of the physical
lattice system - such as the physical
hoppig integrals and orbital energies or certain
external fields - and the $\lambda_i$ are the single-electron
parameters of the reference system which parameterize the self-energy.
We assume that amongst the $\zeta_i$ there is also
a uniform or staggered magnetic field $h$ along the $z$-direction. 
This implies that the values of all single-particle
parameters of the reference system must be taken as spin-dependent:
\begin{eqnarray*}
\lambda_{i,\sigma}= \lambda_{i,+} + sign(\sigma) \lambda_{i,-},
\end{eqnarray*}
which results in a spin-dependent self-energy, $\Sigma_\uparrow(\omega)\ne
\Sigma_\downarrow(\omega)$. 
For a staggered field we switch to the antiferromagnetic unit cell
and assume that the $\lambda_{i,-}$ have opposite
sign at the two  Ni-ions in this cell. This means
that the self-energy for an $\uparrow$-electron
is $\Sigma_\uparrow(\omega)$ at the first Ni-ion and
$\Sigma_\downarrow(\omega)$ at the second Ni-ion in the antiferromagnetic cell
and vice versa for a $\downarrow$-electron (${\bf k}$-sums now
have to be performed over the antiferromagnetic Brillouin zone).
For a uniform field we retain the orginal unit cell
and use the spin dependent self-energy at the single
Ni-ion in this cell.\\
We assume that  we have found a stationary point $\lambda_i^*$
for $h=0$ i.e.:
\begin{equation}
\frac{\partial \Omega}{\partial \lambda_i}\bigg|_{\lambda_i^*}=0
\label{statcon}
\end{equation}
for all $i$ and denote the Grand Potential for this
solution by $\Omega_0$.
Since this is the paramagnetic stationary point
all spin-odd parameters $\lambda_{i,-}$ are zero.
Upon applying a small finite $h$ 
{\em in the lattice system}, $\Omega$ therefore can be 
expanded as
\begin{eqnarray}
\Omega &=&\Omega_0 +  \frac{1}{2}\;\sum_{i,j} \tilde{\lambda}_i \;A_{i,j}\; 
\tilde{\lambda}_j +
\sum_i \tilde{\lambda}_i \;B_i\; h + \frac{1}{2}\; C\;h^2,
\nonumber \\
\label{omexp}
\end{eqnarray}
where the shifts $\tilde{\lambda}_i=\lambda_i- \lambda_i^*$
and $A$, $B$ and $C$ are 2$^{nd}$ derivatives
of $\Omega$ at the point $\lambda_i=\lambda_i^*,h=0$.
There are no terms
linear in the $\tilde{\lambda}_i$ because of (\ref{statcon})
and there is no term linear in $h$ because $\Omega$ must be an even function 
of $h$.
The 2$^{nd}$ derivatives can be evaluated numerically whereby the fact that
the $\lambda_i$ are parameters of the reference system whereas the
staggered field $h$ is one of the $\zeta_i$ in (\ref{Omegafunction})
causes no problem. Moreover, all derivatives are to be evaluated in the
paramagnetic phase, so no calculation in a finite field is necessary.
Demanding stationarity we obtain for the shifts $\tilde{\lambda}_i$:
\begin{eqnarray}
\frac{\partial \Omega}{\partial \tilde{\lambda}_i} &=&
\sum_j\; A_{i,j} \tilde{\lambda}_j + B_{i} h = 0,
\end{eqnarray}
and reinserting into (\ref{omexp}) we obtain $\Omega$
as a function of $h$\cite{vca_lacoo3}
\begin{eqnarray}
\Omega(h)&=&\Omega_0 -\frac{h^2}{2}\;\chi\; \nonumber \\
\chi  &=& \sum_{i,j}\;B_i A_{i,j}^{-1} B_j - C.
\label{chi_exp}
\end{eqnarray}
$\Omega$ must be invariant under a simultaneous
sign change of $h$ and all spin-odd parameters $\bar{\lambda}_{i,-}$
so that $\tilde{\lambda}_{i,+}=0$ for all $i$
and the sums over $i$ and $j$ in (\ref{chi_exp})
extend only over the spin-odd parameters.\\
%%%%%%%%%%%%%%%%%%%%%%%%%%%%%%%%%%%%%%%%%%%%%%%%%%%%%%%%%%%%
\begin{figure}
\includegraphics[width=0.8\columnwidth]{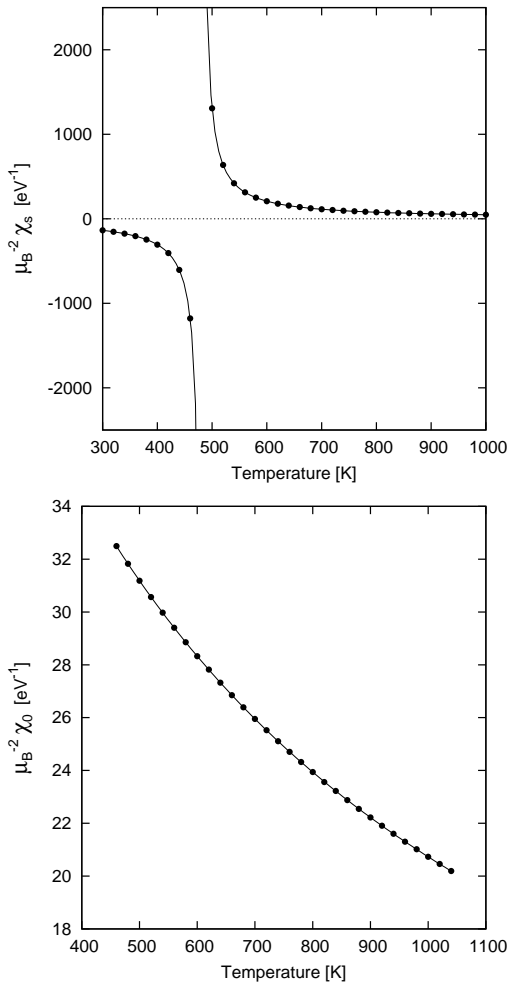}
\caption{\label{fig16} 
Staggered (top) and uniform (bottom) susceptibility of NiO.
The symbols give the calculated values, the lines are the
expressions (\ref{fit1}) and (\ref{fit2}).}
\end{figure}
%%%%%%%%%%%%%%%%%%%%%%%%%%%%%%%%%%%%%%%%%%%%%%%%%%%%%%%%%%%%
Figure \ref{fig16} shows the staggered and uniform 
susceptibilites obtained in this way as a function of temperature.
Thereby the term
\[
H_m = -h\;\sum_i\;e^{i{\bm Q}\cdot{\bm R}_i}\;(n_{i\uparrow}-n_{i\downarrow})
\]
was added to the lattice Hamiltonian, so that the physical
susceptibilities are given by
\[
\chi_{\bm Q}=-\mu_B^2\;\frac{\partial^2 \Omega}{\partial h^2}.
\]
The staggered susceptibility can be fitted accurately by
\begin{equation}
\chi_s(T) = \frac{\mu_B^2\;C_s}{T-T_N}
\label{fit1}
\end{equation}
with $T_N=481.03\;K$ and $C_s=24805\; K\cdot eV^{-1}$
whereas the uniform susceptibility can be fitted by
\begin{equation}
\chi_0(T) = \frac{\mu_B^2\;C_0}{T+\Theta_{CW}}
\label{fit2}
\end{equation}
with $\Theta_{CW}=491.62\;K$ and $C_0=30923\; K\cdot eV^{-1}$.
The divergence of $\chi_s(T)$ at $T_N$ is due to an
eigenvalue of the Hessian $A$ crossing zero at $T_N$.\\
Interestingly the Curie-Weiss temperature
$\Theta_{CW}$ is somewhat higher than the N\'eel temperature
$T_N$. This can be understood as a consequence of the 
type-II antiferromagnetic structure and (weak)
antiferromagnetic exchange between Ni-ions connected
by a $90^o$-degree Ni-O-Ni bond (an example would be two Ni-ions 
at distance $(a,a,0)$). For such a pair of Ni ions
there is a competition between the direct antiferromagnetic
exchange (mediated by the direct Ni-Ni hopping as described
by Slater-Koster parameters such as
$(d d \sigma)$) and the ferromagnetic exchange due to
Hund's rule coupling on oxygen.
Let us assume that the net exchange constant between such
a pair of Ni ions is antiferromagnetic (this is certainly true
for the present calculation which does not include Hund's rule
exchange on oxygen).
Then, any given Ni ion has $12$ neighbors of that type and
in the type-II antiferromagnetic structure the ordered moment
of one half of these neighbors is parallel to the moment of the ion at the 
center whereas it is antiparallel for the other half. The
exchange fields due to these $12$ neighbors therefore cancel and the
N\'eel temperature is determined solely by the
antiferromagnetic superexchange with the
$6$ neighbors connected by $180^o$-degree Ni-O-Ni bonds.
On the other hand the parameter $\Theta_{CW}$ 
is a measure as to how strongly the antiferromagnetic exchange
between spins opposes a uniform ferromagnetic polarization.
For the case of a uniform ferromagnetic polarisation, however,
the exchange fields from all $12$ $(a,a,0)$-like
neighbors are parallel and therefore do contribute
to $\Theta_{CW}$. The discrepancy between $T_N$ and $\Theta$ 
obtained in the VCA calculation thus is to be expected. 
The experimental value $\Theta_{exp}=2000$ K was given in
Ref. \cite{Singer} but being almost four times the experimental
N\'eel temperature this appears somewhat high.\\
For a Heisenberg antiferromagnet the constant $C$ would be given by
\begin{eqnarray*}
C = \frac{S(S+1)}{3k_B}\left(gS_{eff}\right)^2
\end{eqnarray*}
Using the values from the fits in Figure \ref{fig16}
with $S=1$ we obtain the reasonable values
$S_{eff}=1.000$ for $\chi_0$
and $S_{eff}=0.895$ from $\chi_s$.
The smaller value for $S_{eff}$ in the staggered case likely is
due to the fact that in the computation of $\chi_0$
the magnetic field is applied to all orbitals in the unit cell
whereas it acts only on the Ni 3d orbitals in the case of  $\chi_s$.\\
Lastly we point out an interesting feature of these results:
the matrix $A$ of second derivatives of $\Omega$ with respect
to the spin-odd parameters $\lambda_{i,-}$ in (\ref{chi_exp})
obviously is the same in the case of staggered and uniform susceptibility.
Only the quantities $B$ and $C$ are different. Still, the resulting
susceptibilities have a completely different but physically reasonable
temperature dependence.
%%%%%%%%%%%%%%%%%%%%%%%%%%%%%%%%%%%%%%%%%%%%
\section{Antiferromagnetic phase}
%%%%%%%%%%%%%%%%%%%%%%%%%%%%%%%%%%%%%%%%%%%%
To obtain the results presented so far only the paramagnetic 
solution was needed. If we want to discuss the
antiferromagnetic phase itself we need to find stationary points
with spin dependent parameters $\lambda_i$ and this doubling of the
number of parameters complicates the numerical problem of
finding the stationary point. We recall that in the parmagnetic case
a total of $4$ parameters were varied:
$\epsilon_d({e_g})$, $\epsilon_L(e_g)$, $V(e_g)$ and 
$\epsilon_d(t_{2g})$ (moreover, $V(t_{2g})=0$ always
was a stationary point and with that value the last parameter
$\epsilon_L(t_{2g})$ is irrelevant). Doubling all of the
nonvanishing parameters would result 
in a total of $8$ parameters. 
Using the Newton method this is still numerically manageable
but it turned out that a more severe problem appears.
While it might seem that the more parameters $\lambda_i$ one is varying
the better an approximation for the self-energy results,
calculations showed that the opposite is true.
Introducing too many symmetry-breaking parameters $\lambda_{i,-}$ leads 
to unphysical solutions - one example is discussed in detail in Appendix A.
In fact it turned out that retaining more than $2$ - out of the $4$
possible - $\lambda_{i,-}$ leads to unphysical solutions.
Accordingly, in the following we present solutions obtained
with $6$ parameters $\lambda_i$. In choosing the $\lambda_{i,-}$ to be
kept we heuristically use the staggered susceptibility $\chi_s$ as a guidance.
Namely we can restrict the set of the  $\lambda_{i,-}$
in the expression (\ref{chi_exp}) for $\chi_s$
to only $2$ and examine which combination
still gives a $\chi_s$ which is closest
to the one obtained with the full set of $4$ $\lambda_{i,-}$.
It turned out that retaining only the spin-odd part of the
$e_g$-like hopping integral $V_{-}(e_g)$ and the $t_{2g}$-like
$d$-level energy $\epsilon_{d,-}(t_{2g})$ the staggered
susceptibility $\chi_s$ - and in particular the N\'eel temperature -
remain practically unchanged. This appears plausible
because the physical mechanism that stabilizes antiferromagnetism in NiO
is the enhanced hopping for the $e_g$ electrons of one spin direction
along the $180^o$-degree Ni-O-Ni 
bonds connecting sites on different sublattices.
A spin dependent d-level-to-Ligand hopping then clearly is the best way to 
simulate this effect in a cluster with only a single Ni-ion.
Since we have set $V(t_{2g})=0$
in the paramagnetic case, $\epsilon_{d}(t_{2g})$ moreover
is the only remaining parameter pertinent to the $t_{2g}$ orbitals.
In the following, the resulting solution will be referred to as AF-I.\\
In addition there is a second type of antiferromagnetic solution
where $V_{+}(e_g)=V_{-}(e_g)$ so that
the hopping for one spin direction of $e_g$ electrons is exactly
zero. In this case the only remaining spin-odd parameters to be varied
are $\epsilon_{d,-}(t_{2g})$ and $\epsilon_{d,-}(e_{g})$ 
(since one spin direction of
the $e_g$ electrons has zero hopping, the spin splitting of the
$e_g$-like ligand, $\epsilon_{L,-}(e_g)$
is irrelevant and can be set to zero).
It should be stressed that in this case $\Omega$ is stationary also
with respect to $V_{-}(e_g)$, and the values $V_+(t_{2g}) =V_-(t_{2g})=0$ 
remain stationary as in the paramagnetic case. Despite the fact that only 
$6$ parameters are actually varied, 
the corresponding solution therefore is stationary with respect to all
$12$ possible parameters of the cluster. This solution will be
referred to as  AF-II.\\
%%%%%%%%%%%%%%%%%%%%%%%%%%%%%%%%%%%%%%%%%%%%%%%%%%%%%%%%%%%%
\begin{figure}
\includegraphics[width=0.8\columnwidth]{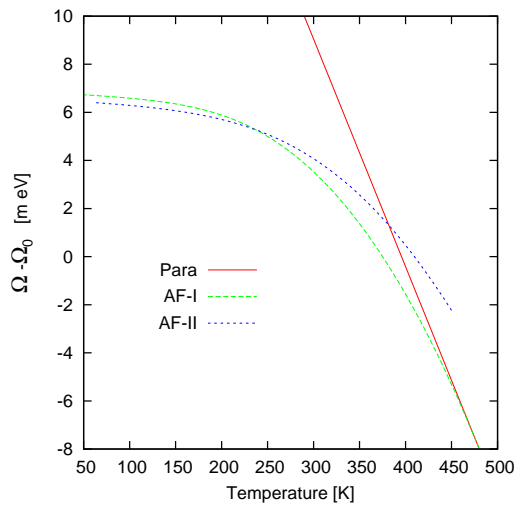}
\caption{\label{fig8} 
Grand Potential per Ni ion
for the various solutions described in the text
as a function of temperature.\\
$\Omega_0=-356.042\;eV$ is an arbitrary reference energy.}
\end{figure}
%%%%%%%%%%%%%%%%%%%%%%%%%%%%%%%%%%%%%%%%%%%%%%%%%%%%%%%%%%%%
Figure \ref{fig8} shows $\Omega$ as a function of
temperature for the  paramagnetic, AF-I and AF-II solutions.
For the paramagnetic phase to good approximation
$\Omega(T)=\Omega_0 - k_B T log(3)$ where $k_B log(3)$ is the
entropy due to the 3-fold degenerate $^3$A$_{2g}$ ground state of a single
Ni$^{2+}$ ion with configuration $t_{2g}^6e_g^2$.
At $T_N$ the solution AF-I branches off as would be expected
for a 2$^{nd}$ order phase transition. At $237.5\;K$ there is a crossing with
finite difference of slopes between the AF-I and the AF-II solutions. 
This would imply a 1$^{st}$ order phase transition which most probably is 
unphysical. Rather, this may be the way in which the VCA
approximates a continuous but rapid change of the electronic state.\\
We discuss the phase transition at $T_N$.
As usual we introduce the staggered magnetization $m_s$
\begin{equation}
m_s = -\frac{\partial \Omega}{\partial h_s}
\label{staggdef}
\end{equation}
and switch to the Legendre-transformed potential
$\Omega'(m_s)=\Omega(h_s) + m_s\cdot h_s$
which - using (\ref{chi_exp}) - is 
\begin{eqnarray*}
\Omega'(m_s)= \frac{1}{2}\;\mu_B^2\;\chi_s^{-1}\;m_s^2 + O(m_s^4).
\end{eqnarray*}
In the present situation where the chemical potential is within
the sizeable insulating gap this equals the Gibbs free energy % $\Phi(m_s)$ 
up to an additive constant.
Comparison with (\ref{fit1}) shows that the lowest order
term in the expansion of $\Omega'(m_s)$ has the form expected
from Landau theory:
\begin{eqnarray*}
\Omega'(m_s) = a(T-T_N) m_s^2 + \frac{b}{2}\;m_s^4
\end{eqnarray*}
with $a=1/(2C_s)$. To carry this further we use
\begin{eqnarray*}
h_s = \frac{\partial \Omega'}{\partial m_s} = 2a(T-T_N) m_s
+ 2bm_s^3.
\end{eqnarray*}
The dependence of $h_s$ on $m_s$ can be obtained within the VCA
by increasing the staggered field $h_s$ in small steps starting
from $h_s=0$, thereby always using the converged AF-I solution for 
the preceding step as starting point for the Newton algorithm
for the next $h_s$. After convergence, $m_s$ is obtained
from (\ref{staggdef}).
%%%%%%%%%%%%%%%%%%%%%%%%%%%%%%%%%%%%%%%%%%%%%%%%%%%%%%%%%%%%
\begin{figure}
\includegraphics[width=0.8\columnwidth]{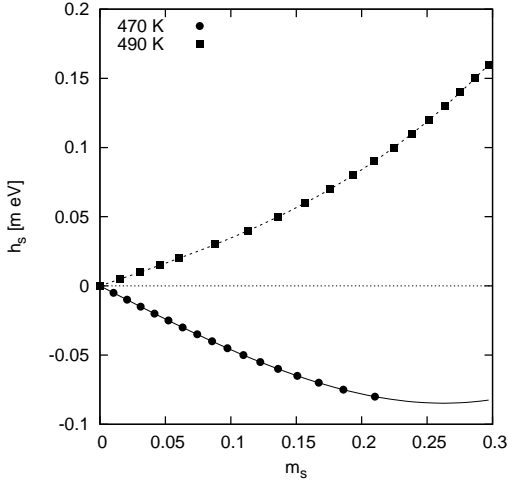}
\caption{\label{fig6} 
Staggered field versus staggered magnetization below and
above $T_N$. The lines are the cubic fits
discussed in the text.}
\end{figure}
%%%%%%%%%%%%%%%%%%%%%%%%%%%%%%%%%%%%%%%%%%%%%%%%%%%%%%%%%%%%
The resulting curves are shown in Figure \ref{fig6}. 
At 470 K the magnetization is opposite to the field because 
below $T_N$ the 
paramagnetic state and hence also the states `close' to it are unstable
and in fact this behaviour is precisely what is expected
from Landau theory. Namely 
$h_s(m_s)$ can be fitted by a third order polynomial
$h_s = c_1 m_s + c_2 m_s^3$ with
$c_1 =-0.484 \;m eV$, $c_2=2.339\; m eV$ at 470 K and
$c_1 =0.322\; m eV$, $c_2=2.458\; m eV$ at 490 K. From the rather 
similar values of
$c_2$ we can conclude that $b \approx 1.20\; m eV$.
Evaluating $c_1$ from $c_1=2a(T-T_N)$
gives $c_1=-0.447\;m eV$ at 470 K and $c_1 =0.362\; m eV$ at 490 K,
reasonably consistent with the values extracted from $m_s(h_s)$.
%%%%%%%%%%%%%%%%%%%%%%%%%%%%%%%%%%%%%%%%%%%%%%%%%%%%%%%%%%%%
\begin{figure}
\includegraphics[width=0.8\columnwidth]{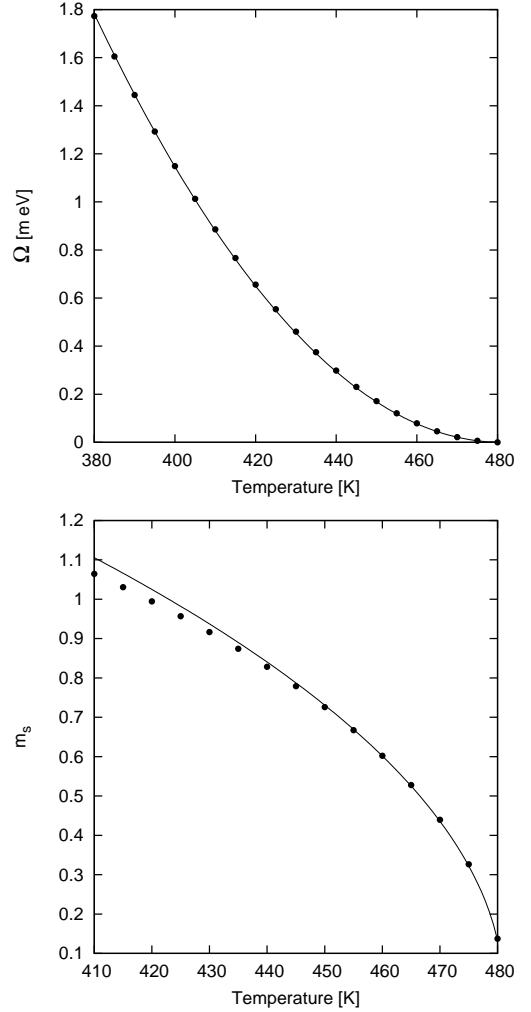}
\caption{\label{fig7} 
Top: Difference of the
Grand Potential in the paramagnetic phase
and in the antiferromagnetic phase versus temperature (dots)
and fit (line).\\
Bottom: Ordered moment versus temperature  (dots)
and fit (line).}
\end{figure}
%%%%%%%%%%%%%%%%%%%%%%%%%%%%%%%%%%%%%%%%%%%%%%%%%%%%%%%%%%%%
Figure \ref{fig7} shows the difference 
$\Omega_{para}- \Omega_{AF}$.
Since both are calculated in zero external field we have
$\Omega'=\Omega$ and we expect
\begin{eqnarray*}
\Omega_{para}(T)-\Omega_{AF}(T) &=& A(T-T_N)^2.
%-\frac{a^2}{2b}(T-T_N)^2.
\end{eqnarray*}
The fit gives $A=1.746\;10^{-4}\;meV\;K^{-2}$ whereas
using $a^2/(2b)=1.694\;10^{-4}\;meV\;K^{-2}$.
Figure \ref{fig7} also shows the ordered moment
$m_s=\langle n_{d,\uparrow}\rangle - \langle n_{d,\downarrow}\rangle$
versus temperature. Close to $T_N$
this can be fitted by $m_s(T)=B\sqrt{T_N-T}$ with
$B=0.131\;K^{-1/2}$ (for comparison: $\sqrt{(a/b)}=0.130\;K^{-1/2}$).
The symmetry-breaking parameters $V_-(e_g)$ and
$\epsilon_{d,-}(e_g)$ have a similar $T$-dependence 
$\propto \sqrt{T_N-T}$ close to $T_N$.
All in all the phase transition is described well
by Landau theory and the VCA allows to extract the
parameters of the theory from the original Hamiltonian.\\
%%%%%%%%%%%%%%%%%%%%%%%%%%%%%%%%%%%%%%%%%%%%%%%%%%%%%%%%%%%%
\begin{figure}
\includegraphics[width=0.8\columnwidth]{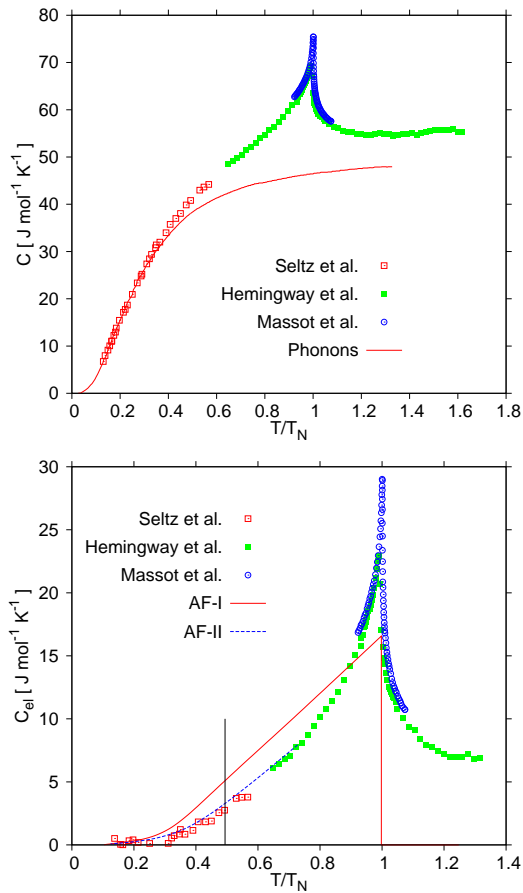}
\caption{\label{fig2} 
Top: Experimental spefific heat of NiO as obtained
by different authors \cite{Seltz,Hemingway,Massot} and phonon 
contribution\cite{Coy}.\\
Bottom: Experimental specific heat after subtraction of the phonon
contribution
compared to the VCA result. The vertical line
gives the temperature where the solutions
AF-I and AF-II cross, see Figure \ref{fig8}.}
\end{figure}
%%%%%%%%%%%%%%%%%%%%%%%%%%%%%%%%%%%%%%%%%%%%%%%%%%%%%%%%%%%%
Next we consider the specific heat $C(T)$ which is shown in
Figure \ref{fig2}.
The top panel shows experimental data for NiO
taken from Refs. \cite{Seltz,Hemingway,Massot}.
Also shown is the phonon contribution which was calculated from the
phonon spectrum measured by inelastic neutron scattering at room 
temperature\cite{Coy}. The electronic heat capacity $C_{el}$
is compared to the
VCA result in the lower part of the figure.
The VCA of course does not reproduce the divergence of
$C_{el}$ at the ordering transition which follows the critical exponents
for a 3D Heisenberg antiferromagnet\cite{Massot}. 
Apart from that the result from the VCA is roughly consistent with the 
measured values,
especially the solution AF-II agrees quite well with the data
at low temperature. In fact, this solution appears to match the experimental 
data considerably better up to $\approx 0.7\;T_N$.
This may be an indication that in NiO this solution is realized
even at temperatures well above the transition between
AF-I and AF-II at $237.5\;K$
(which is indicated by the vertical line in Figure \ref{fig2}).
In fact, as will be seen below, the single
particle spectral function for this solution matches the experimental
photoelectron spectra - which are usually taken at room temperature -
quite well.\\
The apparently large value of $C_{el}$ above
$T_N$ probably is an artefact: the phonon spectrum was measured
at room temperature that means where the lattice was deformed by
magnetostriction. Above $T_N$ this deformation is absent and the phonon spectrum
may change so that using the low temperature phonon spectrum gives 
an incorrect estimate for the phonon contribution.
The electronic heat capacity per mole obeys the sum rule
\[
\int_0^{T_h}\;\frac{C_{el}(T)}{T}\;dT = R\log(3)
\]
where $T_h$ is well above the N\'eel temperature.
It turns out that the experimental data exhaust this
sum rule at $T_h\approx 660\;K$.\\
%%%%%%%%%%%%%%%%%%%%%%%%%%%%%%%%%%%%%%%%%%%%%%%%%%%%%
\begin{table}[h,t]
\begin{center}
\begin{tabular}{|c|rrr|}
\hline
                         &   AF$\;\;\;$&        Para$\;\;\;$& $\Delta$$\;\;\;$  \\
\hline
$ e_d(e_g)\uparrow      $&     -51.4113 &  -51.5278 & \\
$ e_d(e_g)\downarrow    $&     -50.9896 &  -51.5278 & \\
$ V(e_g)  \uparrow      $&      -1.4880 &   -1.7260 & \\
$ V(e_g)  \downarrow    $&       0.0000 &   -1.7260 & \\
$ e_L(e_g)\uparrow      $&       4.3633 &    3.4089 & \\
$ e_L(e_g)\downarrow    $&       4.3633 &    3.4089 & \\
$ e_d(t_{2g})\uparrow   $&     -51.1777 &  -51.2256 & \\
$ e_d(t_{2g})\downarrow $&     -51.0794 &  -51.2256 & \\
\hline
$\Omega+\mu N $         & -244.0363 & -244.0235 &   -0.0128  \\
$\langle H\rangle$      & -244.0344 & -244.0046 &   -0.0299  \\
 $S/k_B$                &    0.1090 &    1.0977 &   -0.9888  \\
$\langle H_0\rangle$    & -457.9924 & -457.3822 &   -0.6102  \\
$\langle H_1\rangle$    &  213.9579 &  213.3776 &    0.5803  \\
$\langle H_{pd}\rangle$  &   -3.5283 &   -3.4380 &   -0.0903  \\
$\langle H_{pp}\rangle$  &   -0.1915 &   -0.1757 &   -0.0158  \\
$\langle H_{dd}\rangle$  &   -0.0030 &   -0.0034 &    0.0004  \\
\hline
$\langle n_{e_g,\uparrow} \rangle$      &    0.2247 &    1.0945 &   -0.8698  \\
$\langle n_{e_g,\downarrow}\rangle$      &     1.9738 &    1.0945 &    0.8793  \\
$\langle n_{d}\rangle$                &    8.1985 &    8.1890 &    0.0095  \\
$\langle n_{t_{2g},\uparrow} \rangle$    &    3.0000 &    3.0000 &    0.0000  \\
$\langle n_{t_{2g},\downarrow} \rangle$   &    3.0000 &    3.0000 &    0.0000  \\
\hline
\end{tabular}
\caption{Top: Parameters of the AF-II (AF) and
paramagnetic (Para) solution at $200$ Kelvin.
$\Delta$=AF-Para.
Center part: Various contributions to $\Omega$ (energies
in $eV$ per Ni)
Bottom part: Occupation numbers of the Ni 3d shell.}
\label{tab2}
\end{center}
\end{table}
%%%%%%%%%%%%%%%%%%%%%%%%%%%%%%%%%%%%%%%%%%%%%%%%%%%%%
Finally, Table \ref{tab2} compares the
parameter values for the paramagnetic and AF-II solutions
at $200$ Kelvin, as well as various observables.
The expectation value of any term $H_{part}$ in the Hamiltonian can be
calculated by replacing $H_{part}\rightarrow \zeta H_{part}$ 
and using
\[
\langle H_{part}\rangle = \frac{\partial \Omega}{\partial \zeta}\bigg|_{\zeta=1}.
\]
The numerical calculation of the derivative thereby is simplified considerably
by taking into account that due to the stationarity condition for
the $\lambda_i$ their variation with $\zeta$ can be neglected\cite{h3}.
The antiferromagnetic unit cell and ${\bf k}$-mesh were used also for the 
paramagnetic calculation to avoid artefacts.\\
A somewhat surpising feature is that the quite different 
cluster parameters for the two different solutions give only slightly different
$\Omega$. 
As expected, $\Omega$ is lower for the antiferromagnetic phase
due to the lower energy $\langle H \rangle$.
This is partly compensated by the almost
complete loss of entropy in the AF phase but at the low
temperature considered this does not result in a higher $\Omega$.
As already mntioned the entropy in the paramagnetic phase 
is close to $S/k_B=log(3)=1.0986$ as expected for a system of localized
$S=1$ spins. While this may seem trivial it should be noted that
the spin degeneracy can only be reproduced if a spin rotation
invariant Hamiltonian is used. Discarding the off-diagonal matrix
elements of the Coulomb interaction (\ref{h1}) breaks the
spin-rotation symmetry so that the entropy of the paramagnetic phase
cannot be obtained correctly.\\
The considerably lower value of $\langle H_0\rangle$ in the
AF phase comes about because the $d$-occupancy increases slightly, 
by $0.0095$.
With the orbital energy $\epsilon_d=-52\;eV$
this lowers $\langle H_0\rangle$ by $-0.494\;eV$. This is more than
compensated, however, by the increase of the Coulomb energy 
$\langle H_1\rangle$ in the AF phase by $0.5803\;eV$.
Eventually, the energy in the antiferromagnetic phase becomes lower
due to the $dp$-hybridization which is lowered by
$-0.0903\;eV$ in the AF phase.
Consistent with the theory of superexchange the driving force
behind the antiferromagnetic ordering is not the
lowering of the Coulomb energy but a gain of kinetic energy.
Interestingly there is also a significant - on the scale
of the change of $\langle H \rangle$ - gain in the direct O 2p-O 2p 
hopping energy $\langle H_{pp}\rangle$ in the AF phase. This is
due to the increased charge transfer to Ni 3d which reduces the filling of the 
O 2p orbitals and thus allows for enhanced O 2p-O 2p hopping. This contribution 
is missed in models for superexchange which consider only a single
Ni-O-Ni bond. \\
To summarize the results of the two preceding sections:
the description of the magnetic properties and phase transition
of NiO as obtained by the VCA is very similar to what would be obtained from 
a simple mean-field treatment of a localized spin system.
It should be noted, however, that the Hamiltonian does
not contain any exchange terms nor is there any molecular field
in the physical system. Rather, the Hamiltonian is the one for the
full NiO lattice  (\ref{h0}), the self-energy is calculated
in a cluster containing a single Ni-ion and the coupling
between Ni-ions is solely due to the lattice kinetic energy.
Still, the VCA even captures subtle details such as the presence
of different exchange channels as manifested by the different values
of $T_N$ and $\Theta_{CW}$. It should also be kept in mind that all parameters 
in the original Hamiltonian are several orders of magnitude larger than the
differences in energy in Table \ref{tab2} but still the calculated 
N\'eel temperature is quite close to the experimental value.
The VCA thus appears successful in correctly extracting the low energy
scales relevant for ordering phenomena and thermodynamics
from the high energy scales of Hubbard $U$, charge transfer energy,
and hopping parameters.
%%%%%%%%%%%%%%%%%%%%%%%%%%%%%%%%%%%%%%%%%%%%%%%%%%%%%%%%%%%
\section{Photoelectron spectra in the antiferromagnetic phase}
%%%%%%%%%%%%%%%%%%%%%%%%%%%%%%%%%%%%%%%%%%%%%%%%%%%%%%%%%%%
We proceed to a discussion of 
`high energy physics' and consider the single particle spectral function.
%%%%%%%%%%%%%%%%%%%%%%%%%%%%%%%%%%%%%%%%%%%%%%%%%%%%%%%%%%%%
\begin{figure}
\includegraphics[width=0.8\columnwidth]{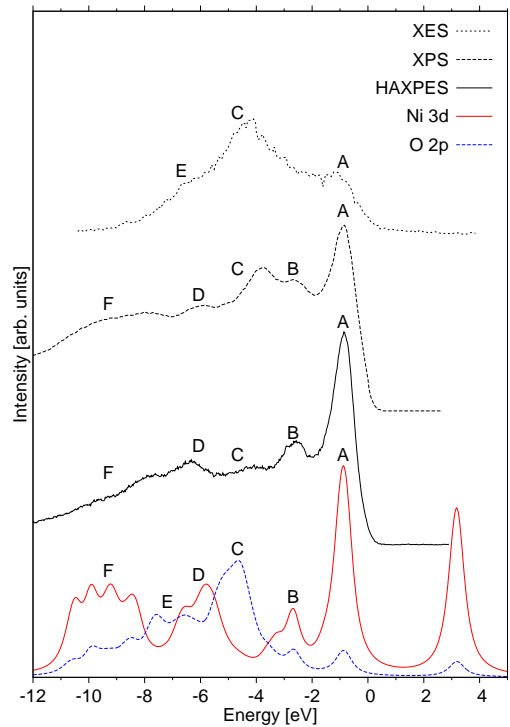}
\caption{\label{fig4}  
$\bm{k}$-integrated spectral function $A(\omega)$ for the antiferromagnetic 
solution AF-II at $200$ Kelvin compared to experimental spectra.}
% %%BoundingBox: -10 40 388 453
\end{figure}
%%%%%%%%%%%%%%%%%%%%%%%%%%%%%%%%%%%%%%%%%%%%%%%%%%%%%%%%%%%%
Figure \ref{fig4} shows the $\bm{k}$-integrated spectral density
\begin{eqnarray}
A(\omega) = -\frac{1}{\pi}\sum_{{\bm k}\alpha} 
G_{\alpha,\alpha}({\bm k},\omega+i\eta),
\label{specden}
\end{eqnarray}
where the sum over $\alpha$ runs over either the Ni 3d or the O 2p orbitals.
This was calculated for the antiferromagnetic solution AF-II at $200$ Kelvin.
Experiments are usually done at room temperature but the specific heat data
in Figure \ref{fig2} suggest that the AF-II solution may be relevant also at 
higher temperature.
The Figure also shows different experimental angle-integrated spectra:
First, hard x-ray photoelectron spectroscopy (HAXPES)
with a photon energy of $h\nu=6500$~eV\cite{Haupricht} - at this energy 
the photoionization cross section for Ni 3d is approximately 10 times larger
than that for O 2p so that predominantly the Ni 3d density of states is observed.
Second, an x-ray photoemission (XPS) spectrum taken with a photon energy
of $h\nu=67$~eV\cite{Tjernberg}. 
This is close to the Ni 3p$\rightarrow$ 3d absorption threshold so that 
the satellite region around $-10\;eV$ is resonantly enhanced\cite{Ohetal}.
And, third, an x-ray emission (XES) spectrum which shows predominantly the
O 2p density of states\cite{Kurmaev}.\\
Several peaks in the theoretical spectra can be identified in 
the various experimental spectra: these are the peaks $A$ and $B$ which have 
Ni 3d character and thus appear in HAXPES and XPS (although
at $h\nu=67$eV the peak $A$ is anti-resonantly suppressed).
The peak $A$ also has some oxygen admixture so that together with peak $C$ 
it can also be seen in the XES spectrum.
Peak $C$ also corresponds to a weak feature observed in HAXPES
and in the XPS spectrum.
Peak $D$ can be seen both in HAXPES and XPS and the tail
of the XES spectrum towards negative energy also shows an indication of the
additional shoulder E which corresponds to a similar feature in the 
theoretical O 2p spectrum. Finally the rather broad feature $F$ can be seen very
well in the XPS spectrum. By and large there is good agreement between 
calculated and measured spectra. It has to be kept in mind, 
however, that as far as the Ni 3d density is concerned
a similar degree of agreement has been
obtained earlier by Fujimori and Minami\cite{c1}
and van Elp {\em et al.}\cite{c7} by the cluster method. The discussion
so far shows mainly that as far as angle-integrated spectra are concerned 
the VCA `inherits' the accuracy of the cluster method.\\
We therefore turn to the quantity which allows for the
most detailed comparison to experiment, namely the ${\bm k}$-resolved
spectral density
\begin{eqnarray}
A({\bm k},\omega) = -\frac{1}{\pi}\sum_{\alpha} 
G_{\alpha,\alpha}({\bm k},\omega+i\eta),
\label{specden1}
\end{eqnarray}
%%%%%%%%%%%%%%%%%%%%%%%%%%%%%%%%%%%%%%%%%%%%%%%%%%%%%%%%%%%%
\begin{figure}
\includegraphics[width=\columnwidth]{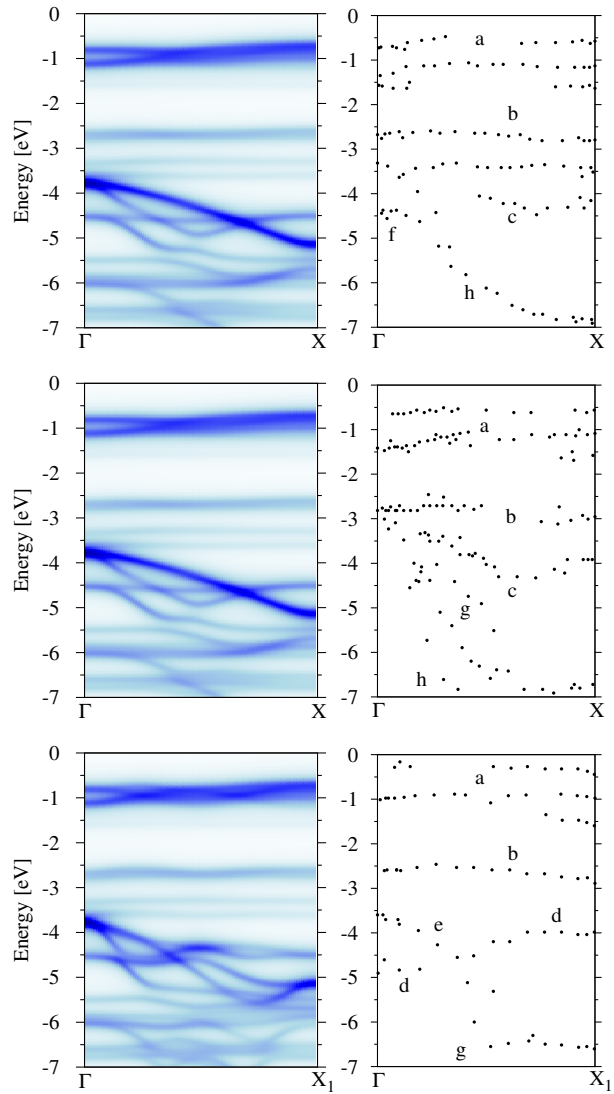}
\caption{\label{fig15}
Single particle spectral density $A({\bm k},\omega)$ 
for the antiferromagneric solution AF-II at $200$ Kelvin compared to
ARPES data by Shen {\em et al.}\cite{Shen_long}. The three panels show the
data for off-normal emission along $\Gamma-X$ (top),
normal emission along $\Gamma-X$ (middle) and
off-normal emission along $\Gamma-X_1$ (bottom).}
\end{figure}
%%%%%%%%%%%%%%%%%%%%%%%%%%%%%%%%%%%%%%%%%%%%%%%%%%%%%%%%%%%%
where the sum over $\alpha$ now runs over both, the Ni 3d and the O 2p orbitals.
The dispersion of peaks in $A({\bm k},\omega)$ can be compared to the band 
structure as measured in angle resolved photoemission spectroscopy (ARPES).
To date there are two ARPES studies of NiO, one  by
Shen {\em et al.}\cite{Shen_long} and the other by 
Kuhlenbeck {\em et al.}\cite{Kuhlenbeck}.
Shen {\em et al.} give three different sets of data points:
the bands from $\Gamma$ to $X=(\frac{2\pi}{a},0,0)$
(i.e. the $(1,0,0)$-direction)  measured
in normal and off-normal emission and the
bands from $\Gamma$ to $X_1=(\frac{2\pi}{a},\frac{2\pi}{a},0)$ 
(i.e. the $(1,1,0)$-direction)) measured
in off-normal emission. The two data sets along $\Gamma-X$ agree for some
bands but differ for others due to
matrix element effects. If a given band is observed
in any one experimental geometry it obviously does exist and if it is not 
observed in another geometry this can only be be a matrix-element effect.
The true band structure along $\Gamma-X$ thus should 
comprise at least the superposition of the two sets of bands for normal 
and off-normal emission.
Figure \ref{fig15} compares $A({\bm k},\omega)$ 
and the respective experimental band dispersions.
For both directions the top of the band structure is formed by a complex 
of several closely spaced bands with high spectral weight
in the range $-0.5\rightarrow -1.5\;eV$, labeled $a$ in the Figure.
In the angle-integrated spectrum in Figure \ref{fig4}
these bands produce the intense peak A.
The high spectral weight of these bands
can also be seen in the experimental spectra in
Figures 7 and 8 of Shen {\em et al.}.
Below this group of bands there is a
gap of approximately $1\;eV$. In the range
$-2.5\rightarrow -3.5\;eV$ there are several essentially dispersionless
bands with weak intensity, labeled $b$. In the angle integrated spectrum
in Figure  \ref{fig4} these bands produce the
weak feature B. Shen {\em et al.} resolved two such bands
along $(1,0,0)$ but only one along $(1,1,0)$ - since the dispersions
must match at $\Gamma$ there probably are
more than one of these dispersionless bands also along $(1,1,0)$.\\
Such (nearly) dispersionless bands can be seen also at even more negative
energies, but there they are superimposed over and mix with the
strongly dispersive O 2p derived bands which results in more
${\bm k}$-dependence.
First, there is the dispersive
band $c$ along $(1,0,0)$ and $e$ along
$(1,1,0)$. In experiment band $c$ shows a relatively strong upward bend
near $X$ - this may indicate that there rather a part of
the dispersionless band which starts at $\Gamma$ at $\approx -4.6\;eV$
has been observed.
This dispersionless band and its `avoided crossing' with a strongly 
dispersive O 2p derived band $c$ as predicted by the
VCA may also have been observed near $\Gamma$, see the 
region labeled $f$ in the top panel.
Along $(1,1,0)$ this dispersionless band can be
followed over the full ${\bf k}$-range, see the
band labeled $d$ in the bottom panel.
In normal emission (middle panel) it moreover becomes apparent that
the strongly downward dispersing O 2p band indeed splits
into two bands with opposite curvature - see the region labeled $g$ - 
which would be similar to the VCA bands. Lastly, at 
$\approx -6.6\;eV$
another dispersionless band - labeled $g$ and $h$ in the middle and
bottom panel -
is observed which would correspond to the nearly dispersionless band
which starts out from $\Gamma$ at slightly below $-6\;eV$ (and which
gives rise to the peak $D$ in the angle-integrated spectrum in Figure
\ref{fig4}).\\
Some of the above interpretations are corroborated in
Figure \ref{fig5} which shows a comparison to the
band structure deduced by Kuhlenbeck {\em et al.}
along $\Gamma-X$\cite{Kuhlenbeck}.
There the dispersionless bands $a$, $b$ and $c$
obviously correspond to the bands with the same labels
in the data by Shen {\em et al.} - and the corresponding bands
predicted by the VCA. Particularly interesting is the band portion 
labeled $d$ in Figure \ref{fig5}
which also shows a peculiar downward curvature and corresponds exactly
to the part labeled $g$ in Figure \ref{fig15} - which in turn had
some counterpart in the VCA bands (plotting the two experimental
band structures on top of each other shows the exact correspondence
of these two bands). Finally, the dispersionless band portion $e$ 
is precisely the continuation towards
$\Gamma$ of the dispersionless band labeled $h$ in Figure \ref{fig15} -
which also has its counterpart in the VCA band strcuture.\\
%%%%%%%%%%%%%%%%%%%%%%%%%%%%%%%%%%%%%%%%%%%%%%%%%%%%%%%%%%%%
\begin{figure}
\includegraphics[width=\columnwidth]{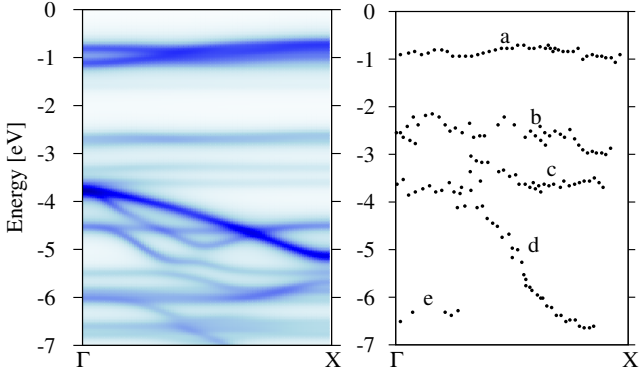}
\caption{\label{fig5}  
Single particle spectral density $A({\bm k},\omega)$ 
for the antiferromagneric solution AF-II at $200$ Kelvin compared to the
ARPES data by Kuhlenbeck {\em et al.}\cite{Kuhlenbeck}.}
\end{figure}
%%%%%%%%%%%%%%%%%%%%%%%%%%%%%%%%%%%%%%%%%%%%%%%%%%%%%%%%%%%%
Generally speaking for all bands which should be
easy to observe because they either have a high intensity or 
are relatively isolated from other bands
there is an essentially one-to-one correspondence between VCA and experiment.
The VCA predicts a multitude of dispersive
low-intensity bands below $\approx -4\;eV$ and only a few of these
seem to have been observed.
Combining the three experimental spectra along $\Gamma-X$ indicates, however,
that the experimental band structure in this energy range
does not consist just of the three strongly dispersive O 2p-derived bands 
obtained by band structure calculations\cite{Mattheiss} which are also
predicted by many `correlated' calculations as well\cite{Kunesetal_band,yinetal}.
Rather, additional bands, both dispersive and non-dispersive,
appear to be observed.
In the next section it will be shown that the dispersionless
bands are in fact the very fingerprints of the atomic multiplets in the ARPES 
spectra.\\
%%%%%%%%%%%%%%%%%%%%%%%%%%%%%%%%%%%%%%%%%%%%%%%%%%%%%%%%%%%%
\begin{figure}
\includegraphics[width=0.8\columnwidth]{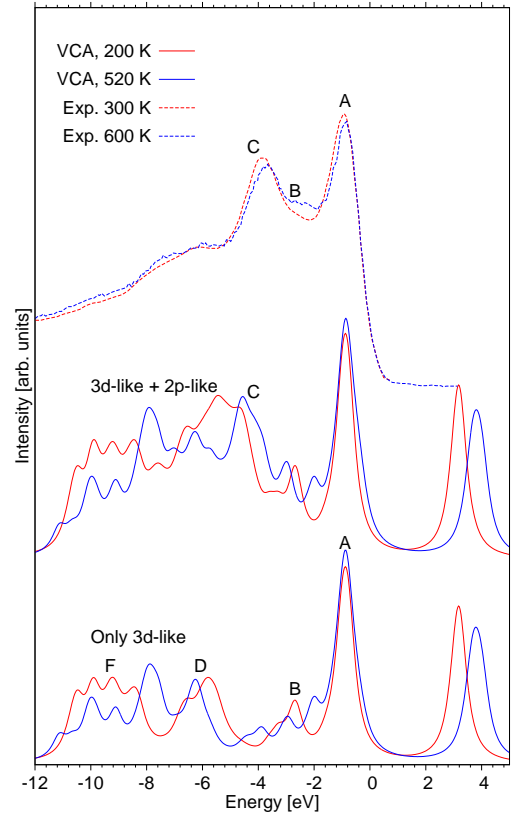}
\caption{\label{fig17}
{\bf k}-integrated spectral density for the antiferromagnetic AF-II
solution at 200 K and the paramagnetic solution at 520 K.
Also shown are experimental spectra by Tjernberg {\em et al.}
measured below and above the experimental N\'eel temperature
of 523 K.}
\end{figure}
%%%%%%%%%%%%%%%%%%%%%%%%%%%%%%%%%%%%%%%%%%%%%%%%%%%%%%%%%%%%
To conclude this section we discuss the temperature dependence of the spectra.
The bottom part of
Figure \ref{fig17} compares the angle integrated Ni 3d-like
spectral function for the AF-II solution at 200 K and for the
paramagnetic solution at 520 K (the paramagnetic solution has practically no
temperature depedence). While the spectra have very similar overall
shape there are small differences. The gap between the large peaks
$A$ and $D$ is filled with weight and the relatively well-defined
peak $B$ more or less disappears in the paramagnetic phase.
A rather strong redistribution of weight occurs in the satellite
region were weight disappears between $-11\;eV$ and $-8\;eV$
and a new strong peak grows at $\approx -8\;eV$.\\
Figure \ref{fig17} also shows the experimental
spectrum taken by Tjernberg {\em et al.}\cite{Tjernberg} at 615 K - 
which is well above $T_N$ - and a modified version of the spectrum at 300 K.
More precisely, the 300 K spectrum was broadened by convolution with
a Gaussian to simulate the enhanced thermal broadening
and fitted to the high-temperature spectrum whereby both
spectra were normalized to unity\cite{Tjernberg}. Due to the 
relatively low photon energy $h\nu=65\:eV$ the experimental spectrum also
contains a considerable amount of O 2p weight, which gives rise to
the intense peak $C$ (compare Figure \ref{fig4}).
Accordingly Figure \ref{fig17}
also shows the sum of Ni 3d-like  and O 2p-like spectral functions.
In the experimental spectrum the peak A looses weight in the paramagnetic phase 
whereas the opposite is predicted by the calculation.
In experiment the spectral weight in the energy range between the two large
peaks $A$ and $C$ increases in the paramagnetic phase
and a similar change occurs in the theoretical spectra were the
relatively well-defined gap between the peaks $A$ and $C$ is partly filled 
in the paramagnetic spectrum, although the effect seems less pronounced
in experiment.In experiment the peak $C$ looses a small amount of
spectral weight and is shifted to slightly less negative energy in the 
paramagnetic phase. A similar tendency can be seen in the theoretical spectra
but considerably exaggerated.
Lastly, in experiment the spectral weight increases slightly at various 
positions in the satellite region below $-8\;eV$
but no decrease is observed anywhere.
In contrast to this in the theoretical spectra there is a drastic change
in the satellite region where a considerable amount of weight disappears 
around $-10\;eV$ and a new strong peak appears at approximately $-8\;eV$.
Summarizing, the VCA is only partly successful in predicting
the changes of the photoemission spectrum across the N\'eel temperature.
It has to be kept in mind, however, that the 
photon energy of $65\;eV$ used in the experiment
is close to the $3p\rightarrow 3d$ absorption
threshold so that the satellite (peak $A$) are resonantly enhanced
(antiresonantly suppressed). In fact, the intensities
of the various peaks are quite different from the HAXPES spectrum in
Figure \ref{fig4}. Accordingly, additional
effects may come into play which determine the intensity of these
features and this might be one explanation why discrepancies with theory
occur precisely for peak $A$ and the satellite. In any way some of the observed
changes with temperature - or actually: between antiferromagnetic and
paramagnetic phase - appear to be reproduced qualitatively by the VCA.
Lastly, Figure \ref{fig17} also shows a somewhat surprising difference 
between the single particle spectra in the antiferromagnetic and paramagnetic 
phase: namely the insulating gap in the paramagnetic phase is larger than
in the antiferromagnetic phase.
More precisely, the peak-to-peak distances are $4.65\;eV$ and $4.05\;eV$
so that the gap increases by $\approx 10\;\%$. So far the temperature 
dependence of the insulating gap in NiO has not been studied experimentally.
As will be discussed in the next section, however, there is a clear physical
reason reason for this discrepancy namely the fact that the mechanism which 
opens the insulating gap in the two phases is quite different.
%%%%%%%%%%%%%%%%%%%%%%%%%%%%%%%%%%%%%%%%%%%%%%%%%%%%%%%%%%%%%
\section{Discussion of the Self Energy}
\begin{figure}
\includegraphics[width=0.8\columnwidth]{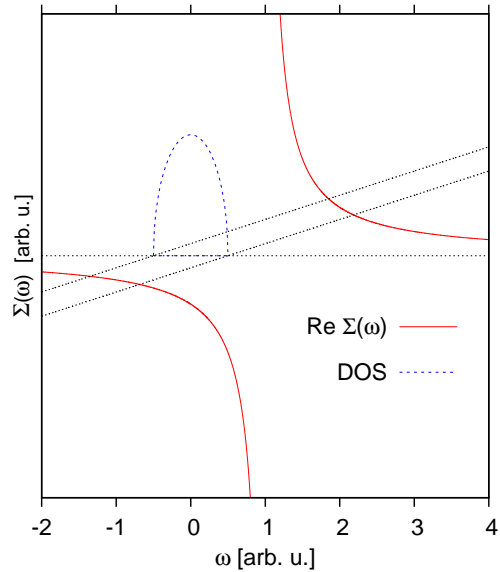}
\caption{\label{fig14}  
Graphical solution of equation (\ref{qpeq}). The self-energy has a single
pole at $\zeta=1$.}
\end{figure}

%%%%%%%%%%%%%%%%%%%%%%%%%%%%%%%%%%%%%%%%%%%%%%%%%%%%%%%%%%%%%
We discuss some of the results presented in
the preceeding section from the `self-energy perspective'.
Luttinger has shown\cite{Luttingerself} that for a single band system
the self-energy has a spectral respresentation of the form
\begin{eqnarray}
\Sigma(\omega) = \eta + \sum_i\;\frac{\sigma_i}{\omega-\zeta_i}
\label{lutty}
\end{eqnarray}
where $\eta$, $\sigma_i>0$ and $\zeta_i$ are real. In the following
we assume these parameters to be ${\bf k}$-independent.
The equation for the poles of the Green's function reads
\begin{equation}
\omega_k -\epsilon_k = \sum_i\;\frac{\sigma_i}{\omega_k-\zeta_i} 
\label{qpeq}
\end{equation}
where for brevity of notation we have replaced
$\epsilon_k + \eta -\mu \rightarrow \epsilon_k$.
Let us first assume that we have only a single pole with a large weight,
Figure \ref{fig14} shows the resulting $\Sigma(\omega)$ 
%%%%%%%%%%%%%%%%%%%%%%%%%%%%%%%%%%%%%%%%%%%%%%%%%%%%%%%%%%%%
%%%%%%%%%%%%%%%%%%%%%%%%%%%%%%%%%%%%%%%%%%%%%%%%%%%%%%%%%%%%
for real $\omega$. Also shown is the noninteracting density of states
for the band $\epsilon_k$, the two straight lines
correspond to $\omega-\epsilon_{-}$ and $\omega-\epsilon_{+}$
where $\epsilon_{-}$ and $\epsilon_{+}$ are the
bottom and top of the noninteracting band $\epsilon_k$.
The intersections of these lines with $\Sigma(\omega)$
give the solutions of the equation (\ref{qpeq}) and there
is one solution for any $k$ in between these. 
A single isolated pole of $\Sigma(\omega)$ thus
splits the noninteracting band into the two Hubbard bands
and opens a gap in the spectral function. Such an isolated
pole with a residuum $\propto N^0$ in the self-energy obviously is the 
very essence of a Mott insulator
- this can also be seen in the self-energy of the
2-dimensional Hubbard model\cite{hubself}.
%%%%%%%%%%%%%%%%%%%%%%%%%%%%%%%%%%%%%%%%%%%%%%%%%%%%%%%%%%%%
\begin{figure}
\includegraphics[width=0.8\columnwidth]{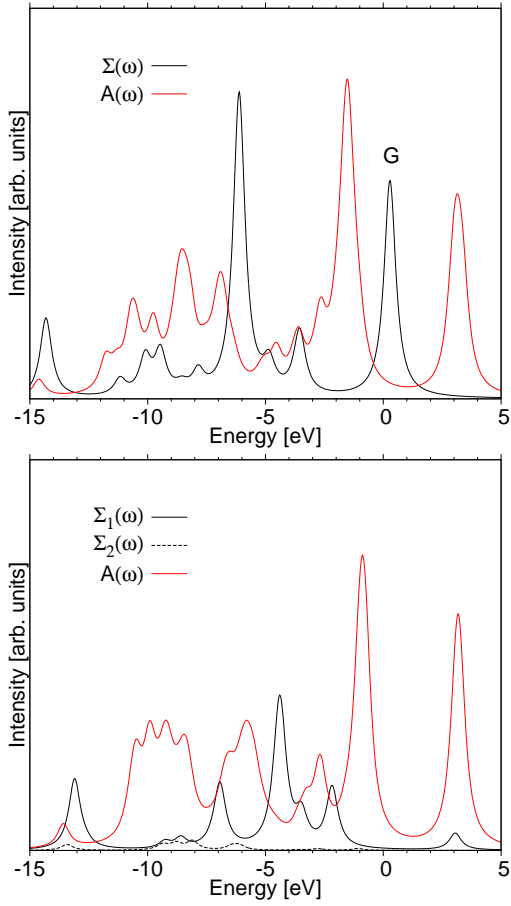}
\caption{\label{fig12}  
Top: $e_g$-like self-energy $\Sigma(\omega)$ and ${\bf k}$-integrated $d$-like
spectral function $A(\omega)$ for the paramagnetic solution at $520$ K.
Bottom: same for the AF-II solution at $200$ K.
The two different self-energies for the antiferromagnetic case refer 
to the two spin-directions. The scale for the self-energy is the same
for both panels.}
\end{figure}
%%%%%%%%%%%%%%%%%%%%%%%%%%%%%%%%%%%%%%%%%%%%%%%%%%%%%%%%%%%%
Figure \ref{fig12} shows the $e_g$-like
self-energy for the paramagnetic solution at $520$ K
and for the antiferromagnetic AF-II solution at $200$ K as well as the 
respective ${\bf k}$-integrated $d$-like spectral function $A(\omega)$.
For the paramagnetic solution there is indeed an isolated intense peak
of the self-energy - labelled $G$ in the Figure - within the insulating gap.
In contrast no such `gap-opening-peak' exists
in the self-energy for the antiferromagnetic solution.
There the mechanism which opens the gap is the different
values of the additive constants $\eta_\uparrow$ and $\eta_\downarrow$ 
in the spin-dependent self-energy, which has the effect of an oscillating 
potential
\[
V_{SDW}(i) = e^{i{\bm Q}\cdot{\bm R}_i}\; \frac{\eta_\uparrow - \eta_\downarrow}{2},
\]
which opens a gap in the same way as in spin-density-wave 
mean-field theory. For the AF-II solution at
200 K $\eta_\uparrow(e_g)=59.89\;eV$ $\eta_\downarrow(e_g)=50.69\;eV$
(whereby the large average $(\eta_\uparrow+\eta_\downarrow)/2\approx 55.3\;eV$
cancels the double-counting correction to the
Ni 3d-level energy).
Since the insulating gaps in the paramagnetic and antiferromagnetic 
phase are created by different mechanisms it is not too
surprising that they have different values (see Figure \ref{fig17}). 
%%%%%%%%%%%%%%%%%%%%%%%%%%%%%%%%%%%%%%%%%%%%%%%%%%%%%%%%%%%%
\begin{figure}
\includegraphics[width=0.8\columnwidth]{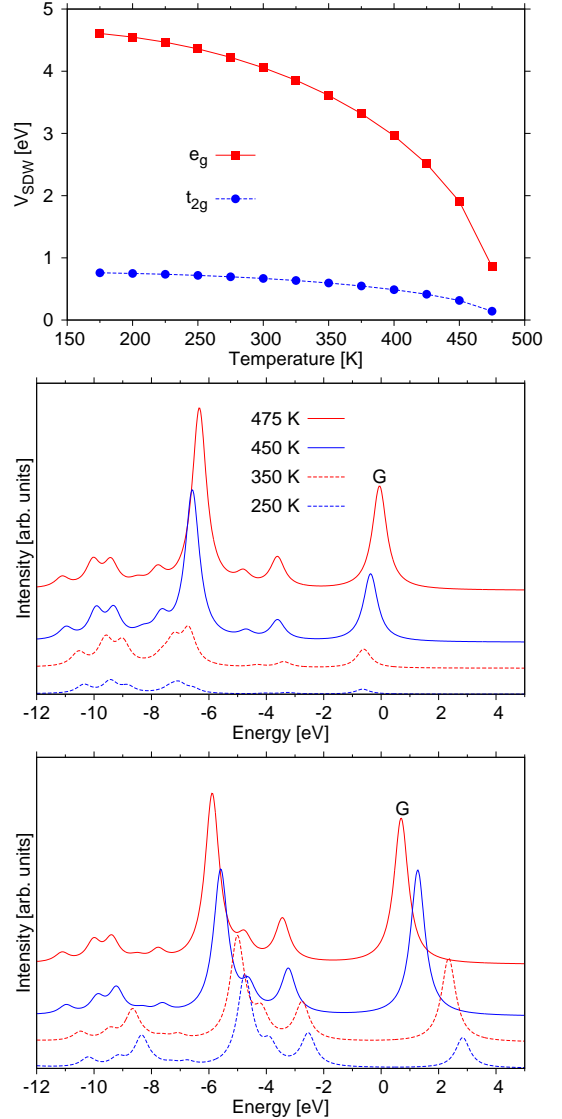}
\caption{\label{fig13} 
Temperature variation of $V_{SDW}=(\eta_\uparrow-\eta_\downarrow)/2$
and the $e_g$-like self-energy
for the two spin directions in the AF-I solution.}
\end{figure}
%%%%%%%%%%%%%%%%%%%%%%%%%%%%%%%%%%%%%%%%%%%%%%%%%%%%%%%%%%%%
For completeness we mention that the $t_{2g}$-like self-energy has no such 
`gap-opening-peak' in either the paramagnetic or insulating phase.
This is special for NiO because  in the $^3A_{2g}$
ground state of d$^8$ in cubic symmetry, which is $t_{2g}^6 e_g^2$,
the $t_{2g}$ orbitals are completely filled and thus comprise a
`band-insulating' subsystem. Very probably this is also the
reason why $V(t_{2g})=0$ is a stationary point.\\
The AF-I solution which branches off the paramagnetic solution at $T_N$
and crosses with the AF-II solution at $237.5\;K$
(see Figure \ref{fig8}) interpolates between these two types of insulating 
gap: Figure \ref{fig13} shows $\Sigma(\omega)$ for this solution
for different temperatures. With decreasing temperature
the peak $G$ which opens the insulating gap decreases and shifts
to the lower or upper edge of the gap whereas
the value of $\eta_\uparrow - \eta_\downarrow$ increases. NiO thus changes smoothly
from a Mott-insulator to a spin-density-wave insulator.\\
%%%%%%%%%%%%%%%%%%%%%%%%%%%%%%%%%%%%%%%%%%%%%%%%%%%%%%%%%%%%
\begin{figure}
\includegraphics[width=\columnwidth]{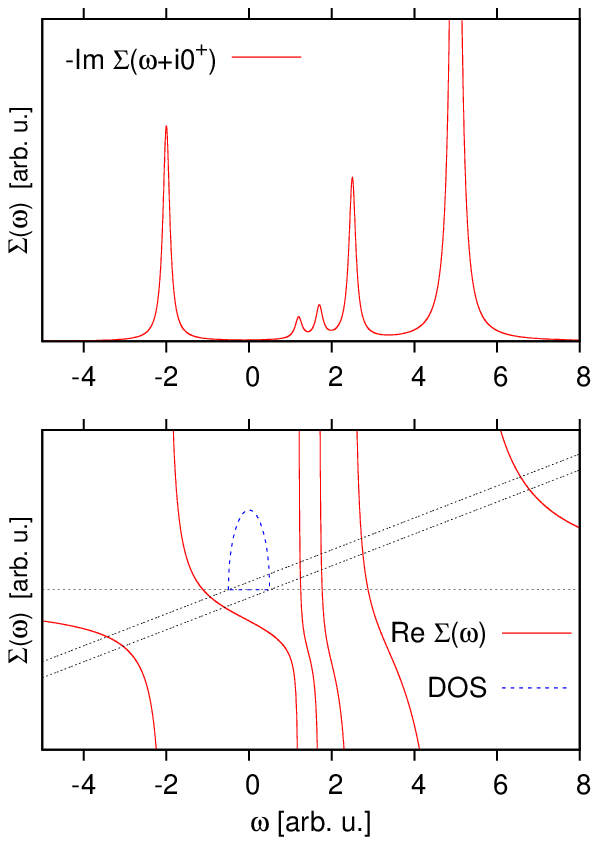}
\caption{\label{fig9}
Top: Spectral density of the `model self-energy' $\Sigma(\omega)$.\\
Bottom: Graphical solution of equation (\ref{qpeq}) for
the quasiparticle energies $\omega_k$ with
the `model self-energy' $\Sigma(\omega)$.
}
\end{figure}
%%%%%%%%%%%%%%%%%%%%%%%%%%%%%%%%%%%%%%%%%%%%%%%%%%%%%%%%%%%%
Next, we discuss the origin of the dispersionless bands observed
in the ARPES spectra of both Shen {\em et al.}\cite{Shen_long}
and Kuhlenbeck  {\em et al.}\cite{Kuhlenbeck}. For the sake of illustration
we consider a `model self-energy' obtained by arbitrarily choosing a few
$\sigma_i$ and $\zeta_i$ in (\ref{lutty}).
Figure \ref{fig9} shows the imaginary part of the resulting
$\Sigma(\omega)$ for $\omega$ slightly off the real axis (top) and 
$\Sigma(\omega)$ for  real $\omega$ (bottom).
The bottom part again shows the density of states of the noninteracting
band $\epsilon_k$ as well as the lines
$\omega-\epsilon_-$ and $\omega-\epsilon_+$.
For real $\omega$ $\Sigma(\omega)$ takes any value in 
$[-\infty,\infty]$ precisely once in any interval $[\zeta_i,\zeta_{i+1}]$
so that the line $\omega-\epsilon_k$ intersects $\Sigma(\omega)$ once for each
$\epsilon_k$. The point of intersection thereby
is between between those of the lines  $\omega-\epsilon_{-}$ and 
$\omega-\epsilon_{+}$. This shows that in between any two successive
poles of the self-energy there is one complete quasiparticle band.
If a given pole has a small $\sigma_i$, however,
$\Sigma(\omega)$ drops almost vertically near the corresponding
$\zeta_i$, so that the width of the respective band becomes
small. Replacing
\[
\sum_i\;\frac{\sigma_i}{\omega-\zeta_i} \rightarrow
C + \frac{\sigma_0}{\omega-\zeta_0}
\]
in the neighborhood of such a pole $\zeta_0$, the resulting
dispersion and quasiparticle weight
$Z=\left(1-\frac{\partial \Sigma}{\partial \omega}\right)^{-1}$ 
are
\begin{eqnarray*}
\omega_k &\approx& \zeta_0 + \frac{\sigma_0}{\zeta_0-C-\epsilon_k},
\nonumber \\
Z_k &\approx& \frac{\sigma_0}{(\zeta_0-C-\epsilon_k)^2}.
\end{eqnarray*}
Therefore, unless the denominator happens to cross zero near $\zeta_0$
this results in a band with little dispersion and low
spectral weight close to $\zeta_0$.
Whether the band is on the high or low energy side of
$\zeta_0$ depends on the sign of the denominator $\zeta_0-C-\epsilon_k$.
Figure \ref{fig11} compares partial ARPES spectra along
$(1,0,0)$, where the sum
in (\ref{specden1}) is restricted to either $e_g$-like or $t_{2g}$-like
Ni 3d orbitals and the respective self-energies.
Although the situation in NiO is more complicated due to the multi-band
situation and the hybridization with the O 2p bands
it is quite obvious how the various dispersionless bands
can be associated with poles of the self-energy.
In the case of NiO these poles describe the multiplet splitting
of the final state of the photoemission process, i.e.
mainly the Ni$^{3+}$ ion. In the absence of the Coulomb
interaction a single Ni 3d shell would have eigenstates obtained by
distributing the electrons over the $e_g$ and $t_{2g}$ levels and the
single particle spectral function $A(\omega)$ would have few peaks
corresponding to the energies of these CEF levels.
The considerably larger number of CEF-split multiplet states 
in the presence of Coulomb interaction - as given e.g. in the
Tanabe-Sugano diagrams -
increases the number of peaks in $A(\omega)$ and the 
interacting peak structure
is generated by the poles of the self-energy of the Ni 3d
electrons in exactly the same way as in Figure \ref{fig9}.
In the solid these poles of the self-energy then generate the
dispersionless bands observed in ARPES as discussed above.
In that sense one can literally see the
dispersionless self-energy of the Ni 3d electrons
directly in the experimental data of Shen {\em et al.} and
Kuhlenbeck {\em et al.}.
%%%%%%%%%%%%%%%%%%%%%%%%%%%%%%%%%%%%%%%%%%%%%%%%%%%%%%%%%%%%
\begin{figure}
\includegraphics[width=\columnwidth]{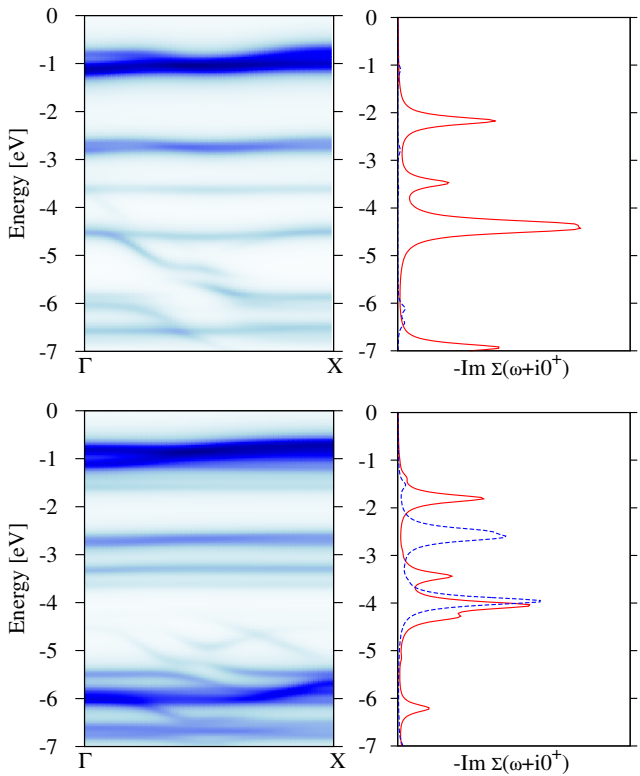}
\caption{\label{fig11} 
Comparison of partial ARPES spectra and respective self-energy
for $e_g$ orbitals (top) and $t_{2g}$ orbitals (bottom).
The two different self-energies refer to the two spin directions.
}
\end{figure}
%%%%%%%%%%%%%%%%%%%%%%%%%%%%%%%%%%%%%%%%%%%%%%%%%%%%%%%%%%%%
%%%%%%%%%%%%%%%%%%%%%%%%%%%%%%%%%%%%%%%%%%%%5
%%%%%%%%%%%%%%%%%%%%%%%%%%%%%%%%%%%%%%%%%%%%5
\section{Conclusion}
%%%%%%%%%%%%%%%%%%%%%%%%%%%%%%%%%%%%%%%%%%%%5
In summary, the Variational Cluster Approximation proposed by
Potthoff allows to combine the classic field theroretical work
of Luttinger and Ward with the very successful cluster method
due to Fujimori and Minami resulting in
an efficient band structure method for strongly correlated
electron systems. Since the VCA is based on exact diagonalization
which is free from the minus-sign problem
it allows to take into account the full Coulomb interaction
in the TM 3d-shell which is known to be crucial for reproducing
the correct multiplet structure and for obtaining agreement with experiment
for angle-integrated valence band 
photoemission\cite{c2,c3,c4,c5,c6,c7,c8,c9,c10} 
and X-ray absorption \cite{x0,x1,x2,x3,x4,x5,x6,x7,x8,xr1,xr2}. 
As might have been expected on the basis of the success of the cluster 
method in describing these spectroscopies, the multiplet structure turns 
out to be important also for reproducing the experimental valence band 
structure as observed in ARPES in that it produces a number of nearly
dispersionless bands observed there. The VCA moreover delivers 
an estimate for the Grand Potential and - as demonstrated above - allows the 
discussion of thermodynamics and phase transitions. It thereby gives a 
unified description for a wide variety of experimental quantities which probe 
energy scales from the $meV$ range up to $\approx 10\;eV$.\\
In the case of NiO, using realistic values of the Hubbard-$U$ and charge 
transfer energy $\Delta$ - as demonstrated by the position of the satellite 
and  the magnitude of the insulating gap - and a moderately adjusted 
value of the Slater-Koster parameter $(pd\sigma)$
(increased by $10\%$ as compared to the LDA band structure estimate)
the theoretical N\'eel temperature is $481$ Kelvin
(experimental value:  $523$ Kelvin). The behaviour near $T_N$
is consistent with a 2$^{nd}$ order phase transition in a local-moment
system, with quite accurate Landau behaviour of the free energy 
and ordered moment below $T_N$ and a Curie-Weiss susceptibility above $T_N$. 
Consistent with experiment the angle-integrated density of states is very
similar for the paramagnetic and antiferromagnetic phase. The angle-integrated 
spectrum and band structure in the antiferromagnetic phase agree 
well with experiment, whereby the band structure shows a considerable number 
of both dispersive and dispersionless bands and again
shows the massive impact of the strong correlations in NiO
in that it differs strongly from the band structure obtained within DFT.\\
Acknowledgement: The author would like to thank F. Hardy and K. Grube
for help with the specific heat data and R. Heid for providing
the ${\bf k}$-meshs for Brillouin zone integration.
%%%%%%%%%%%%%%%%%%%%%%%%%%%%%%%%%%%%%%%%%%%%
\section{Appendix A}
\begin{figure}
\includegraphics[width=0.8\columnwidth]{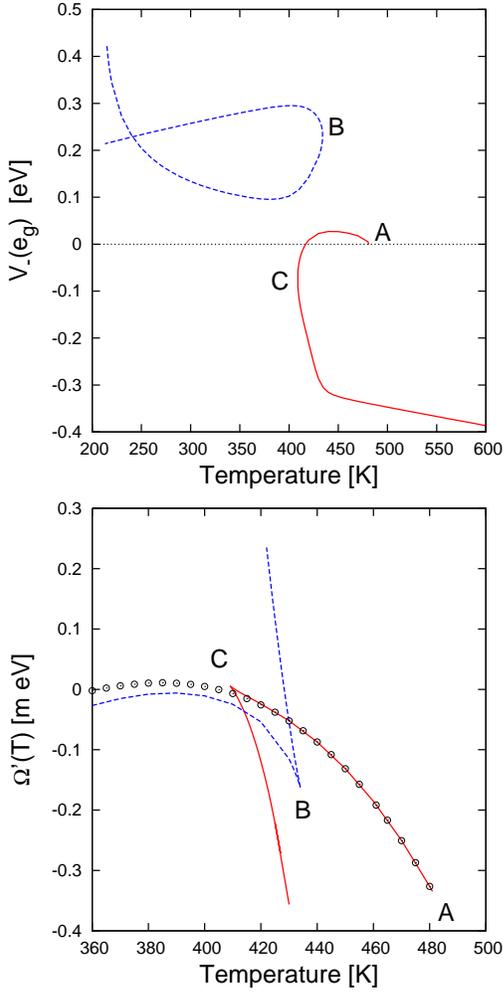}
\caption{\label{fig18} 
Top: Spin-odd hopping parameter $V_{-}(e_g)$ for the solution with
$7$ parameters as a function of temperature.
Bottom: Grand Potential as a function of temperature
for the different solutions with $7$ parameters (lines).
To make small changes visible the function
$f(T)=-2.99067e\cdot10^{-7}T^2 + 1.06805\cdot10^{-4}T -712.08$
was subtracted from $\Omega$. Also shown is $\Omega'(T)$ for
the solution AF-I with $6$ parameters (circles).}
\end{figure}

\begin{figure}
\includegraphics[width=0.8\columnwidth]{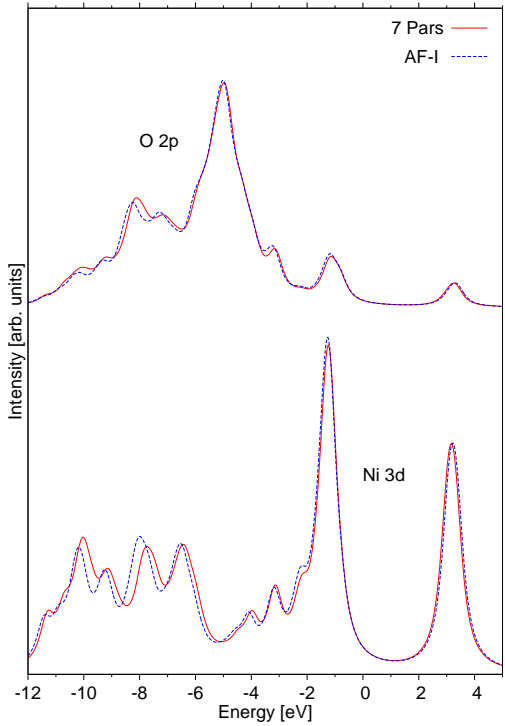}
\caption{\label{fig1} 
Comparison of the  ${\bf k}$-integrated spectral densities
for the AF-I solution with $6$ varied parameters
and the solution with $7$ varied parameters at $400$ K.}
\end{figure}

%%%%%%%%%%%%%%%%%%%%%%%%%%%%%%%%%%%%%%%%%%%%
In this Appendix we discuss an unphysical solution which
appears when $7$ parameters are varied.
Thereby all $4$ spin-even paramaters were varied
and in addition to the spin-odd parameters
$V_{-}(e_g)$ and $\epsilon_{e,-}(t_{2g})$ which are used in the AF-I
solution, also the spin-odd part of the $e_g$ level energy,
$\epsilon_{e,-}(e_g)$. The upper part of Figure \ref{fig18} shows the 
%%%%%%%%%%%%%%%%%%%%%%%%%%%%%%%%%%%%%%%%%%%%%%%%%%%%%%%%%%%%
%%%%%%%%%%%%%%%%%%%%%%%%%%%%%%%%%%%%%%%%%%%%%%%%%%%%%%%%%%%%
temperature dependence of the spin-odd d-Level-to-Ligand hopping
$V_{-}(e_g)$. We could have chosen any other spin-odd
parameter but this one is sufficient to discuss what is happening.
The lower part of the Figure shows the $\Omega'(T)=\Omega(T) - f(T)$
where $f(T)$ is a second order polynomial which has no
physical significance and was subtracted
from $\Omega(T)$ for the sole purpose of making tiny variations 
of $\Omega$ around this relatively strongly varying but smooth 
`background' visible.\\
As one would expect, $V_{-}(e_g)$ starts to deviate from zero
at the N\'eel temperature $T_N=481$ K, see point $A$ in the 
upper part of Figure \ref{fig18}.
At $T_1\approx 434$ K a bifurcation occurs and two
new solutions appear, see point $B$. At $T_2\approx 409$ K
another bifurcation occurs and the solution starting out from $A$
disappears (point $C$). The second solution emerging from the 
bifurcation $C$ can be followed up to temperatures far above $T_N$.
As can be seen in the bottom part, the solution along $A\rightarrow C$ has 
a higher $\Omega$ than the one extending
from $C$ to high temperatures (the part above $430$ K is
omitted for this solution in the bottom part of Figure \ref{fig18}
to keep the range of $\Omega$ sufficiently small). In fact, 
$\Omega$ for this solution turns out to be
even lower than for the paramagnetic solution so that we have
an antiferromagnetic solution which gives the
lowest $\Omega$ up to the highest temperatures studied.
Moreover, the magnetic solution  $A\rightarrow C$ appearing at $T_N$
- which would be consistent with the staggered susceptibility -
would be an unstable state which never should be
realized. One of the two solutions emerging from the bifurcation $B$ 
intersects the unphysical antiferromagnetic solution
at $\approx 420$ K. \\
The bottom part also shows $\Omega'(T)$ for
the solution AF-I with $6$ varied parameters.
Over almost the entire temperature range this solution is very close in 
energy (the deviation is $\propto 10^{-5}\;eV$ between
$300$ K and $420$ K and even smaller above $440$ K) to one of
the solutions with $7$ parameters. In fact, it seems to
interpolate between two branches of  solutions with $7$ parameters.
%%%%%%%%%%%%%%%%%%%%%%%%%%%%%%%%%%%%%%%%%%%%%%%%%%%%%%%%%%%%%%%%%%%%%%%%%%%
\begin{table}[h,t]
\begin{center}
\begin{tabular}{|c|rr|}
\hline
                         &   AF-I$\;\;\;$&        7  Pars\\
\hline
$V_{-}(e_g)$            &  0.0338 &  0.1021  \\
$e_{d,-}(e_g)$          &  0.0000 &  0.0414  \\
$e_{d,-}(t_{2g}$        &  0.0171 &  0.0198  \\
\hline
$\Omega+\mu N $         & -244.043562 &   -244.043570 \\ 
$\langle H\rangle$      &  244.0169   &    244.0169   \\ 	  
 $S/k_B$                &    0.7734   &       0.7736  \\ 	  
$\langle H_0\rangle$    & -457.6312   &    -457.6322  \\ 	  
$\langle H_1\rangle$    &  213.6143   &     213.6153  \\ 	  
$\langle H_{pd}\rangle$ &   -3.4749   &      -3.4750  \\ 	  
$\langle H_{pp}\rangle$ &   -0.1822   &      -0.1822  \\ 	  
$\langle H_{dd}\rangle$ &   -0.0032   &      -0.0032  \\     
\hline
$\langle n_{e_g,\uparrow} \rangle$   &   0.5325	&   0.5328  \\ 
$\langle n_{e_g,\downarrow}\rangle$  &   1.6604	&   1.6601  \\ 
$\langle n_{d}\rangle$               &   8.1928 &   8.1929 \\  
$\langle m_s\rangle$                 &  -1.1279	&  -1.1273  \\ 
\hline
\end{tabular}
\caption{Comparison of the AF-I solution with $6$ parameters
and the solution for $7$ parameters at $400$ K. All energies
in $eV$.}
\label{tab3}
\end{center}
\end{table}
%%%%%%%%%%%%%%%%%%%%%%%%%%%%%%%%%%%%%%%%%%%%%%%%%%%%%%%%%%%%%%%%%%%%%%%%%%%%
That the `closeness' is not restricted to $\Omega$ can be seen
in Table \ref{tab3} which compares some physical quantities
for the two different solutions, and from Figure \ref{fig1}
which compares the ${\bf k}$-integrated spectral densities.
Both, the table and the Figure, shows an essentially perfect agreement
between the two solutions as far as observable quantities are
concerned. In contrast to this, the symmetry-breaking
parameters $\lambda_{i,-}$ which are also listed in the table
are quite different and in fact substantially larger for the solution
with $7$ parameters. This shows that the symmetry-breaking effects of 
$V_{-}(e_g)$ and $e_{d,-}(e_g)$ must cancel to a large degree
in order to simulate the effect of a significantly
smaller $V_{-}(e_g)$ alone. Adding $e_{d,-}(e_g)$ to the set of parameters to 
be varied is superfluous. 
For completeness we note that keeping $e_{d,-}(t_{2g})=0$
so that $V_{-}(e_g)$ remains as
the only spin-odd parameter to be varied, gives results which are 
almost identical to those for the solution AF-I.
From the above it looks very much as if already with $6$ 
parameters the solution is converged with respect to the number of parameters 
and that adding an additional spin-odd parameter results in no more 
significant changes to obervables but creates a new branch of unphysical 
solutions.\\
%%%%%%%%%%%%%%%%%%%%%%%%%%%%%%%%%%%%%%%%%%%%%%%%%%%%%%%%%%%%
%%%%%%%%%%%%%%%%%%%%%%%%%%%%%%%%%%%%%%%%%%%%%%%%%%%%%%%%%%%%
We now discuss the artificial antiferromagnetic solution at high 
temperatures. Table \ref{tab4} compares some
observables for this solution and the paramagnetic one
at $600$ K.
As already mentioned, the unphysical solution has lower $\Omega$
than the paramagnetic phase and in addition also
a higher entropy. There are no
large differences in the various ground state expectation values.
Although the symmetry breaking
parameters $V_{-}(e_g)$ and $e_{d,-}(e_g)$ are substantially larger
than those
for the antiferromagnetic solutions at $400$ K given in Table \ref{tab3}
the ordered moment is much smaller. This shows again that the symmetry 
breaking effect of the different spin-odd parameters cancels almost 
completely in this solution.\\
The above example shows that including too many symmetry breaking parameters
into the subset of variational parameters can lead to unphysical solutions.
Here it is interesting to note that Kozik {\em et al.} recently
reported unphysical solutions obtained by DMFT
for the self-energy of a simple Hubbard-dimer\cite{Koziketal}.
A tendency to produce additional unphysical solutions may be a 
general feature of schemes which aim at computing the self-energy.
Within the VCA the simplest solution to this problem would be to simply 
restrict the number of parameters to a minimum - as was done in the 
solution AF-II in the main text. Clearly this introduces a certain
arbitraryness regarding the subset of parameters to be varied.
On the other hand the above discussion shows that results of
the VCA appear to converge rather well with the number of 
variational parameters. In any way it seems desirable to
find criteria which allow to identify unphysical solutions
or methods to regularize the variational procedure so that
unphysical solutions are suppressed.
%%%%%%%%%%%%%%%%%%%%%%%%%%%%%%%%%%%%%%%%%%%%%%%%%%%%%%%%%%%%%%
\begin{table}[h,t]
\begin{center}
\begin{tabular}{|c|rr|}
\hline
                         &   AF $\;\;\;$&        Para\\
\hline
$V_{-}(e_g)$            &  -0.3865  &   0.0000 \\
$e_{d,-}(e_g)$          &  -0.2424  &   0.0000 \\
$e_{d,-}(t_{2g}$        &  -0.0126  &   0.0000 \\
\hline
$\Omega+\mu N $         &  244.0621 & -244.0613 \\
$\langle H\rangle$      & -244.0046 & -244.0045 \\
 $S/k_B$                &    1.1139 &    1.0981 \\
$\langle H_0\rangle$    & -457.3758 & -457.3815 \\
$\langle H_1\rangle$    &  213.3712 &  213.3770 \\
$\langle H_{pd}\rangle$ &   -3.4373 &   -3.4379 \\
$\langle H_{pp}\rangle$ &   -0.1756 &   -0.1757 \\
$\langle H_{dd}\rangle$ &   -0.0034 &   -0.0034 \\
\hline
$\langle n_{e_g,\uparrow} \rangle$   & 1.1138 & 1.0945  \\
$\langle n_{e_g,\downarrow}\rangle$  & 1.0757 & 1.0945  \\
$\langle n_{d}\rangle$               & 8.1889 & 8.1890  \\
$\langle m_s\rangle$                 & 0.0376 & 0.0000  \\
\hline
\end{tabular}
\caption{Comparison of the unphysical AF solution
and the paramagnetic solution at $600$ K. All energies
in $eV$.}
\label{tab4}
\end{center}
\end{table}
%%%%%%%%%%%%%%%%%%%%%%%%%%%%%%%%%%%%%%%%%%%%%%


\begin{thebibliography}{}
%%%%%%%%%%%%%%%%%%%%%%%%%%%%%%%%%%%%%%%%%%%%%%
\bibitem{deBoer} 
J. H. de Boer and E. J. W. Verwey, Proc. Phys. Soc. London {\bf 49}, 59 (1937).
\bibitem{Mattheiss}
L. F. Mattheiss, Phys. Rev. B {\bf 5}, 290 (1972).
\bibitem{Terakura}
K. Terakura, T. Oguchi, A. R. Williams and J. Kubler, 
Phys. Rev. B {\bf 30}, 4734 (1984).
\bibitem{SawatzkyAllen}
G. A. Sawatzky and J. W. Allen, 
Phys. Rev. Lett.  {\bf 53}, 2339 (1984).
\bibitem{NormanFreeman}
M. R. Norman and A. J. Freeman,
Phys. Rev. B {\bf 33}, R8896 (1986).
%%%%%%%%%%%%%%% NiO Theory
\bibitem{SvaGu} % SIC
A. Svane and O. Gunnarsson, Phys. Rev. Lett. {\bf 65}, 1148 (1990).
\bibitem{Szotek} % SIC
Z. Szotek, W. M. Temmerman, and H. Winter,
Phys. Rev. B 47, 4029(R) (1993).
\bibitem{Czyzyk} %LDA+U,
V. I. Anisimov, I. V. Solovyev, M. A. Korotin, M. T. Czyzyk, 
and G. A. Sawatzky, Phys. Rev. B {\bf 48}, 16929 (1993).
\bibitem{Bengone} %LDA+U, AF NiO
O. Bengone, M. Alouani, P. Bl\"ochl, and J. Hugel,
Phys. Rev. B {\bf 62}, 16392 (2000).
\bibitem{Ary} % GW
F. Aryasetiawan and O. Gunnarsson,
Phys. Rev. Lett. {\bf 74}, 3221 (1995). 
\bibitem{Massi} % GW
S. Massidda, A. Continenza, M. Posternak, and A. Baldereschi
Phys. Rev. B {\bf 55}, 13494 (1997).
\bibitem{Kobayashi_etal} %LDA+U+GW
S. Kobayashi, Y. Nohara, S. Yamamoto, and T. Fujiwara,
Phys. Rev. B {\bf 78}, 155112 (2008).
\bibitem{Manghi}  % 3-body scattering, para
F. Manghi, C. Calandra, and S. Ossicini,
Phys. Rev. Lett. {\bf 73}, 3129 (1994).  
\bibitem{Iga} % 3-body scattering, AF
M. Takahashi and J. I. Igarashi,
Phys. Rev. B {\bf 54}, 13566 (1996);
M. Takahashi and J. I. Igarashi, Ann. Phys. {\bf 5}, 247 (1996).
\bibitem{Kunesetal}% DMFT, DOS
 J. Kunes, V. I. Anisimov, A. V. Lukoyanov, and D. Vollhardt,
Phys. Rev. B {\bf 75}, 165115 (2007).
\bibitem{Kunesetal_band} %DMFT, band structure
J. Kunes, V. I. Anisimov, S. L. Skornyakov, A. V. Lukoyanov, and
D. Vollhardt, Phys. Rev. Lett. {\bf 99}, 156404 (2007).
\bibitem{yinetal} %DMFT, para
Q. Yin, A. Gordienko, X. Wan, and S. Y. Savrasov
Phys. Rev. Lett. {\bf 100}, 066406 (2008).
\bibitem{MiuraFujiwara} %DMFT, AF-phase
O. Miura and T. Fujiwara,
Phys. Rev. B {\bf 77}, 195124 (2008) .
\bibitem{GillenRobertson} %DFT, improved  exc functional
R. Gillen and J. Robertson,
J. Phys.: Condens. Matter {\bf 25}, 165502 (2013).
%%%%%%%%%%%%%Cluster method, XPS %%%%%%%%%%%%%%%%%%%%%
\bibitem{c1} % {FujimoriMinami}
A. Fujimori and F. Minami, Phys. Rev. B {\bf 30}, 957 (1984).
\bibitem{c2} % {Fujimori_Fe2O3}
A. Fujimori, M. Saeki, N. Kimizuka, M. Taniguchi, and S. Suga,
Phys. Rev. B {\bf 34}, 7318 (1986).
\bibitem{c3} % {Ghijsen_CuO}
J. Ghijsen, L. H. Tjeng, J. van Elp, H. Eskes, J. Westerink, 
G. A. Sawatzky, and M. T. Czyzyk, Phys. Rev. B {\bf 38}, 11322 (1988).
\bibitem{c4} % {Fujimori_MnO}
A. Fujimori, N. Kimizuka, T. Akahane, T. Chiba, S. Kimura, F. Minami, 
K. Siratori, M. Taniguchi, S. Ogawa and S. Suga, 
Phys. Rev. B {\bf 42}, 7580 (1990).
\bibitem{c5} % {Elp_MnO}
J. van Elp, R. H. Potze, H. Eskes, R. Berger, and G. A. Sawatzky, 
Phys. Rev. B {\bf 44}, 1530 (1991). 
\bibitem{c6} % {Elp_CoO}
J. van Elp, J. L. Wieland, H. Eskes, P. Kuiper, G. A. Sawatzky, 
F. M. F. de Groot, and T. S. Turner, Phys. Rev. B {\bf 44}, 6090
 (1991).
\bibitem{c7} % {Elp}
J. van Elp, H. Eskes, P. Kuiper, and G. A. Sawatzky,
Phys. Rev. B {\bf 45}, 1612 (1992).
\bibitem{c8} % {Abbate_LaCoO3}
M. Abbate, R. Potze, G. A. Sawatzky, and A. Fujimori,
Phys. Rev. B 49, 7210 (1994).
\bibitem{c9} % {Saitho_LaCoO3}
T. Saitoh, T. Mizokawa, and A. Fujimori, M. Abbate, Y. Takeda, and M. Takano,
Phys. Rev. B {\bf 55}, 4257 (1997).
\bibitem{c10} % {Taguchi_etal}
M. Taguchi, M. Matsunami, Y. Ishida, R. Eguchi, A. Chainani, Y. Takata, 
M. Yabashi, K. Tamasaku, Y. Nishino, T. Ishikawa, Y. Senba, H. Ohashi, 
and S. Shin, Phys. Rev. Lett. {\bf 100}, 206401 (2008).
%%%%%%%%%%%%%%% Cluster method XAS %%%%%%%%%%%%%
\bibitem{x0} 
G. van der Laan, J. Zaanen, G. A. Sawatzky, R. Karnatak, and J.-M. Esteva,
Phys. Rev. B {\bf 33}, 4253 (1986).
\bibitem{x1} % XAS an Molekühlen
G. van der Laan, B. T. Thole, G. A. Sawatzky, and M. Verdaguer,
Phys. Rev. B  {\bf 37}, 6587(R) (1988).
\bibitem{x2} %CoO
K. Okada and A. Kotani,
J. Phys. Soc. Jpn.  {\bf 61}, 449 (1992).
\bibitem{x3} %CoO
A. Tanaka and T. Jo,
J. Phys. Soc. Jpn. {\bf 61}, 2040  (1992).
\bibitem{x4} % {deGroot_94}
F. M. F. de Groot, Journal of Electron Spectroscopy and Related Phenomena,
{\bf 67} 529 (1994).
\bibitem{x5} %{Pen}
H. F. Pen, L. H. Tjeng, E. Pellegrin, F. M. F. de Groot, G. A. Sawatzky, 
M. A. van Veenendaal, and C. T. Chen, Phys.~Rev.~B {\bf 55}, 15500 (1997).
\bibitem{x6} %{Finazzi}
M. Finazzi, N. B. Brookes, and F. M. F. de Groot,
Phys.~Rev.~B {\bf 59}, 9933 (1999).
\bibitem{x7} %{Haverkort_LaCoO3}
M. W. Haverkort, Z. Hu, J. C. Cezar, T. Burnus, H. Hartmann, 
 M. Reuther, C. Zobel, T. Lorenz, A. Tanaka, N. B. Brookes,
 H. H. Hsieh, H.-J. Lin, C. T. Chen, and L. H. Tjeng
 Phys. Rev. Lett. {\bf 97}, 176405 (2006).
\bibitem{x8} %{Merz}
M. Merz, D. Fuchs, A. Assmann, S. Uebe, H. v. L\"ohneysen, P. Nagel,
and S. Schuppler,
Phys.~Rev.~B {\bf 84}, 014436 (2011).
\bibitem{xr1} %{deGroot_05}
F. M. F. de Groot, Coordination Chemistry Reviews, {\bf 249} 31 (2005).
\bibitem{xr2} %{deGroot}
 F. M. F. de Groot and A. Kotani: {\em Core Level Spectroscopy of Solids}
  (Taylor And Francis, 2008).
%%%%%%%%%%%%%%% Cluster method Core level
%\bibitem{Westra}
%J. Zaanen, C. Westra, and G. A. Sawatzky,
%Phys. Rev. B {\bf 33}, 8060 (1986).
%\bibitem{Core_levels}
%A. E. Bocquet, T. Mizokawa, T. Saitoh, H. Namatame, and A. Fujimori,
%Phys. Rev. B {\bf 46}, 3771 (1992).
%%%%%%%%%%%%%%%%% Multiplet Theory
\bibitem{Slater}
J. C. Slater, {\em Quantum Theory of Atomic Structure}, McGraw-Hill,
(1960).
\bibitem{Griffith}
J. S. Griffith, {\em The Theory of Transition-Metal Ions},
Cambridge University Press, 1964.
\bibitem{Sugano}
S. Sugano, Y. Tanabe, and H. Kamimura,
{\em Multiplets of Transition Metal Ions in Crystals}
(Academic Press New York, 1970).
\bibitem{Kanamori}
J. Kanamori, Prog. Theo. Phys. {\bf 30}, 275 (1963).
\bibitem{Marel}
D. van der Marel and G. A. Sawatzky, Phys. Rev. B {\bf 37}, 10674 (1988 ).
\bibitem{maurits_thesis}
M. W. Haverkort, PhD. Thesis University of Cologne (2005);
see also arXiv:cond-mat/0505214. 
\bibitem{p1}
M. Potthoff, Eur. Phys. J. B {\bf 32}, 429 (2003).
\bibitem{p2}
M. Potthoff, Eur. Phys. J. B {\bf 36}, 335 (2003).
\bibitem{p3}
M. Potthoff, M. Aichhorn, and C. Dahnken,
Phys. Rev. Lett. {\bf 91}, 206402 (2003).
\bibitem{vca_1}
R. Eder, Phys. Rev. B {\bf 78}, 115111 (2008).
%%%%%%%%%%%%%%%%%%%%% VCA-refs
\bibitem{LuttingerWard}
J. M. Luttinger and J. C. Ward,
Phys. Rev.{\bf  118}, 1417 (1960).
\bibitem{Nonperturbative}
M. Potthoff, Condens. Mat. Phys. {\bf 9}, 557 (2006).
\bibitem{review}
M.~Potthoff in "Theoretical Methods for Strongly Correlated Systems", 
edited by A. Avella and F. Mancini, Springer (2011);
see also preprint arXiv:11082183.
%%%%%%%%%%%%%%%%%%%%%%%%%%%%%%%%%%%%%%%%%%%%%%%%%%%%%
\bibitem{h1} %Antiferromagnet in 2D Hubbard
C. Dahnken, M. Aichhorn, W. Hanke, E. Arrigoni, and M. Potthoff, 
Phys. Rev. B {\bf 70}, 245110 (2004).
\bibitem{h2} %Hubbard
D. S\'en\'echal, P.-L. Lavertu, M.-A. Marois, and A.-M. S. Tremblay,
Phys. Rev. Lett. {\bf 94}, 156404 (2005).
\bibitem{h3} %phase sep, af, supercon, 2D Hubbard, T=0
M. Aichhorn, E. Arrigoni, M. Potthoff, and W. Hanke, 
Phys. Rev. B {\bf 74},  235117 (2006).
\bibitem{h4} %Hubbard
A. H. Nevidomskyy, C. Scheiber, D. S\'en\'echal, and A.-M. S. Tremblay
Phys. Rev. B {\bf 77}, 064427 (2008).
\bibitem{h5} %Hubbard
M. Balzer, B. Kyung, D. S\'en\'echal, A.-M. S. Tremblay, and M. Potthoff,
Europhys. Lett. {\bf 85}, 17002 (2009).
\bibitem{h6} %Hubbard ferro
M. Balzer and M. Potthoff,
Phys. Rev. B {\bf 82}, 174441 (2010).
\bibitem{h7}
K. Seki, R. Eder, and Y. Ohta,
Phys. Rev. B {\bf 84}, 245106 (2011).
\bibitem{h8}
A. Yamada, K. Seki, R. Eder, and Y. Ohta,
Phys. Rev. B {\bf 88}, 075114 (2013).
\bibitem{h9}
A. Yamada, Phys. Rev. B {\bf 89}, 195108 (2014).
\bibitem{h10}
A. Yamada, Phys. Rev. B {\bf 90}, 235138 (2014).
\bibitem{h11}
A. Yamada, Phys. Rev. B {\bf 90}, 245139 (2014).
\bibitem{r1} %CrO_2
L. Chioncel, H. Allmaier, E. Arrigoni, A. Yamasaki, M. Daghofer, M. I.
  Katsnelson, and A. I. Lichtenstein, Phys. Rev. B {\bf 75}, 140406 (2007).
\bibitem{r2} %TiN
H. Allmaier, L. Chioncel, and E. Arrigoni, 
Phys. Rev. B {\bf 79}, 235126 (2009).
\bibitem{r3} %TiOCl
M. Aichhorn, T. Saha-Dasgupta, R. Valenti, S. Glawion, M. Sing, and
R. Claessen, 
Phys. Rev. B {\bf 80}, 115129 (2009).
\bibitem{vca_lacoo3}
R. Eder, Phys. Rev. B {\bf 81}, 035101 (2010).
\bibitem{r4} %NiMnSb
H. Allmaier, L. Chioncel, E. Arrigoni, M. I. Katsnelson, and A. I.
Lichtenstein, 
Phys. Rev. B {\bf 81}, 054422 (2010).
\bibitem{b1} %Bosons
W. Koller and N. Dupuis, 
J. Phys.: Condens. Matter {\bf 18}, 9525 (2005).
\bibitem{b2} %Bosons
M. Knap, E. Arrigoni, and W. von der Linden, 
Phys. Rev. B {\bf 81}, 235122 (2010).
\bibitem{Singer}
J. R. Singer, Phys. Rev. 929, {\bf 104}, (1956).
\bibitem{Seltz}
H. Seltz, B. J. DeWitt, and H. J. McDonald,
Am. Chem. Soc. {\bf 62} 88 (1940). 
\bibitem{Hemingway}
B. S. Hemingway, American Mineralogist {\bf 75}, 781 (1990).
\bibitem{Massot}
M. Massot, A. Oleaga, A. Salazar, D. Prabhakaran, M. Martin, 
P. Berthet, and G. Dhalenne,
Phys. Rev. B {\bf 77}, 134438 (2008).
\bibitem{Coy} %Room temperature phonon DOS
R. A. Coy, C. W. Thompson, and E. G\"urmen,
Solid State Commun. {\bf 18}, 845 (1976).
\bibitem{Haupricht}
%HAXPES
T. Haupricht, J. Weinen, A. Tanaka, R. Gierth, S. G. Altendorf, 
Y.-Y. Chin, T. Willers, J. Gegner, H. Fujiwara, F. Strigari, 
A. Hendricks, D. Regesch, Z. Hu, Hua Wu, K.-D. Tsuei, Y. F. Liao,
H. H. Hsieh,  H.-J. Lin, C. T. Chen, and L. H. Tjeng,
preprint arXiv:1210.6675.
\bibitem{Tjernberg}
%Finite temperature and Resonant Spektrum
O. Tjernberg, S. S\"oderholm, G. Chiaia, R. Girard,
U. O. Karlsson, H. Nyl\'en, and I. Lindau, Phys. Rev. B {\bf 54}, 10245 (1996).
\bibitem{Ohetal}%Resonant XPS
 S.-J. Oh,J. W. Allen, I. Lindau, and J. C. Mikkelsen, Jr., 
Phys. Rev. B {\bf 26}, 4845 (1982).
\bibitem{Kurmaev}
%XES-Data
E. Z. Kurmaev, R. G. Wilks, A. Moewes, L. D. Finkelstein, S. N. Shamin, 
and J. Kunes,
Phys. Rev. B {\bf 77}, 165127 (2008).
\bibitem{Shen_long}
%Shen-ARPES
Z. X. Shen, R. S. List, D. S. Dessau, B. O. Wells, O. Jepsen,
A. J. Arko, R. Barttlet, C. K. Shih, F. Parmigiani, J. C. Huang, 
and P. A. P. Lindberg, 
Phys. Rev. B {\bf 44}, 3604 (1991).
\bibitem{Kuhlenbeck}
H. Kuhlenbeck, G. Od\"orfer, R. Jaeger, G. Illing, M. Menges, Th. Mull, 
H.-J. Freund, M. P\"ohlchen, V. Staemmler, S. Witzel, C. Scharfschwerdt, 
K. Wennemann, T. Liedtke, and M. Neumann,
Phys. Rev. B {\bf 43} (1991 ).
\bibitem{Luttingerself}
J. M. Luttinger, Phys. Rev. {\bf 121}, 942 (1961).
\bibitem{hubself}
R. Eder, K. Seki, and Y. Ohta
Phys. Rev. B {\bf 83}, 205137 (2011)
\bibitem{Koziketal}
E. Kozik, M. Ferrero, and A. Georges, arXiv:1407.5687.
\end{thebibliography}
\end{document}